\documentstyle[epsf,astrcite]{mn}
\newcommand{\un}[2]{\mbox{\rm\thinspace #1$^{#2}$}}

\newcommand{\be}[1]{\begin{equation}\label{#1}}
\newcommand{\ee}{\end{equation}}

\newcommand{\Eq}[1]{Eq.\,(\ref{#1})}

\newcommand{\Fig}[1]{Fig.\,\ref{#1}}

\newcommand{\gsim}{\mathrel{\hbox{\rlap{\lower.55ex \hbox {$\sim$}}
                   \kern-.3em \raise.4ex \hbox{$>$}}}}
\newcommand{\lsim}{\mathrel{\hbox{\rlap{\lower.55ex \hbox {$\sim$}}
                   \kern-.3em \raise.4ex \hbox{$<$}}}}
\newcommand{\msun}{\mbox{M$_\odot$}}
\newcommand{\pder}[2]{\frac{\partial #1}{\partial #2}}

\newcommand{\Sect}[1]{Sect.\,\ref{#1}}

\newcommand{\spr}[1]{^{\rm #1}}
\newcommand{\sub}[1]{_{\rm #1}}
\newcommand{\Tab}[1]{Table\,\ref{#1}}

\newcommand{\tder}[2]{\frac{{\rm d} #1}{{\rm d} #2}}

\newcommand{\pd}[2]{\frac{\partial #1}{\partial #2}}

\newlength{\ffh}

\newcommand{\bL}{\bmath{L}}
\newcommand{\bcL}{\bmath{\cal L}}
\newcommand{\bl}{\bmath{\ell}}
\newcommand{\noval}{\multicolumn{2}{c}{~}}
\title[Warped accretion discs in X-ray binaries]
    {Warped accretion discs and the long periods in X-ray binaries}
\author[R. A. M. J. Wijers and J. E. Pringle]
       {Ralph A. M. J. Wijers$^{1,2}$ and J. E. Pringle$^1$\\
        $^1$Institute of Astronomy, Madingley Road, Cambridge CB3 0HA, UK\\
	$^2$Dept.\ of Physics and Astronomy, SUNY at Stony Brook,
	    Stony Brook, NY\,11794-3800, USA\\
	E-mail: {\tt rwijers@astro.sunysb.edu} and {\tt jep@ast.cam.ac.uk}}
\date{\underline{submitted to MNRAS, 17-Dec-97, revised submit 2-Nov-98}}

\begin{document}

\maketitle

\begin{abstract}
Precessing accretion discs have long been suggested as explanations
for the long periods observed in a variety of X-ray binaries, most
notably Her\,X-1/HZ Her.  We show that an instability of the disc's
response to the radiation reaction force from the illumination by the
central source can cause the disc to tilt out of the orbital plane and
precess in something like the required manner. The rate of precession
and disc tilt obtained for realistic values of system parameters
compare favourably with the known body of data on X-ray binaries with
long periods.  We explore other possible types of behaviour than
steadily precessing discs that might be observable in systems with
somewhat different parameters. At high luminosities, the inner disc tilts
through more than 90 degrees, i.e.\ it rotates counter to the usual direction,
which may explain the torque reversals in systems such as
4U\,1626$-$67.
\end{abstract}


%%%%%%%%%%%%%%%%%%%%%%%%%%
% body of paper
%%%%%%%%%%%%%%%%%%%%%%%%%%

%%%%%%%%%%%%%%%%%%%%%%%%%%%%%%%%%%%%%%%%%%%%%%%%%%%%%%%%%%%%%%%%%%%%%%%%%%%%%%%%
   \section{Introduction}
   \label{intro}
%%%%%%%%%%%%%%%%%%%%%%%%%%%%%%%%%%%%%%%%%%%%%%%%%%%%%%%%%%%%%%%%%%%%%%%%%%%%%%%%

Soon after the discovery of Her\,X-1 (Tananbaum et~al.\
1972)\nocite{tgkls:72} it was found that beside the 1.24-sec pulse
period and the 1.7-day orbital period (marked by eclipses of the
neutron star by the companion) the system displayed another (`long' or
`third') period: a 35-day cycle of X-ray on and off
states. Photometric variations showed a period of 1.6 days, the beat
period between the orbital and long periods. A phenomenological
`clockwork' model in which the system contains a retrogradely
precessing disc was found to explain the photometry quite well. The
photometric variations are due to the varying amount of area of the
disc facing us, as well as variations in the fraction of the star that
is eclipsed when the disc passes in front of it and the variable X-ray
illumination of the star (Gerend \& Boynton 1976, Deeter et~al.\
1976)\nocite{gb:76,dcgb:76}. The behaviour is not strictly periodic:
there are `anomalous' on- and off-states in X rays, and the period of
the cycle varies by 5--10\% over time (\"Ogelman et al.\
1985)\nocite{okptv:85}.  With time, two more systems were found with
nearly strictly periodic long variations: SS\,433 has jets which
precess around the sky every 164 days, and therefore presumably so
does the underlying disc, and the direction of precession is
retrograde (see Margon 1984)\nocite{margo:84}.  LMC\,X-4 has a
30.4-day long period, which resembles that of Her\,X-1 in many ways,
except that the amplitude is much lower due to the fact that the
companion star is much brighter (Heemskerk et~al.\
1994)\nocite{hp:94}.

The success of the clockwork model quickly led to the general
acceptance of the basic hypothesis that the long periods are due to
precessing tilted accretion discs (Katz 1973, Roberts
1974)\nocite{katz:73,rober:74}.  It appears easy to make a disc in a
binary precess, since the companion star exerts forces on the disc
which, when averaged over the orbit, lead to a constant rate of
retrograde precession of a tilted annulus in the disc.  The problem is
to understand how the disc as a whole can precess at one rate, since
the precession frequency is a strong function of radius in the disc,
and how it remains tilted, i.e.\ avoids sinking back into the
orbital plane on a viscous time scale. Roberts proposed that the disc
is slaved to a companion whose spin is misaligned with the orbit
(Roberts 1974). But that is difficult on general grounds: all neutron
star binaries have eccentric orbits soon after the neutron star has
formed, due to the effects of the supernova explosion. But now most
(notably Her\,X-1) have orbits that are accurately circular,
indicating that tidal forces have been strong enough to damp this
initial eccentricity. It is hard to see how they would at the same
time not have largely damped out a spin-orbit misalignment (Petterson
1977)\nocite{pette:77}, since the spin angular momentum is much less
than the orbital angular momentum for such binary systems.

Beside the three very regular and clear third periods mentioned above,
there are many systems where a less well-defined variability exists.
Some are quite regular but of more uncertain nature, e.g.\ Cyg\,X-1
(294\,d; Priedhorsky, Terrell \& Holt 1983\nocite{pth:83}) and
4U\,1820$-$30 (176\,d; Priedhorsky \& Terrell 1984a)\nocite{pt:84}. Others
are simply less regular, such as Cen\,X-3 (120--160\,d; Priedhorsky \&
Terrell 1983)\nocite{pt:83}. In this latter category numerous objects have
recently been added with well-sampled light curves obtained with the Rossi
X-ray Timing Explorer (RXTE).  Cyg\,X-2 (78\,d; Wijnands, Kuulkers \&
Smale 1996)\nocite{wks:96} and X\,2127$+$119 (37\,d; Corbet, Peele \&
Smith 1997)\nocite{cps:97} have periods that seem fairly stable, but the
amplitude and light curve shape are variable so the periodicity cannot
always be seen equally well. SMC\,X-1, on the other hand, has a fairly
stable amplitude but a period that slowly decreases from over 60\,d to
under 50\,d in the first 600 days of RXTE data (Wojdowski
et~al.\ 1997)\nocite{wclwz:97}.

Moreover, there appears to be some evidence of discs precessing
progradely. First, there has been a detection of an 11.2-day period in
the light curve of Cyg\,X-2 (Holt et~al.\ 1976)\nocite{hbsk:76}.  This
is the beat period between the 78-day long period and the 8.9-day
orbital period, but {\it only if\/} the disc precesses
progradely. Note that this cannot be due to tidally forced
precession. In X\,1916$-$053, the optical period is 0.9\% longer than
the X-ray period, and if the beat period of 3.8 days between the two
is interpreted as the period of disc precession, it would most easily
be for a progradely precessing disc.

It was noted by Petterson (1977)\nocite{pette:77}
and Iping \& Petterson (1990)\nocite{ip:90} that a sufficiently strong
illumination from the centre could cause a disc to maintain a warped,
tilted shape. While probably qualitatively correct, their results
suffer from the use of an incorrect equation of disc evolution (see 
Papaloizou \& Pringle 1983)\nocite{pp:83}.
An accretion disc is indeed unstable to tilting and warping due to radiation
reaction forces when the luminosity of its central source exceeds a 
critical value (Pringle 1996, 1997)\nocite{pring:96,pring:97}. We examine
here the consequences of this instability for the behaviour of discs
in X-ray binaries. We show that the instability provides a mechanism
for sustaining a tilted disc and making it precess with a period that
agrees well with the observed long periods in X-ray binaries such as the
famous 35-day period of Her\,X-1. They also provide the possibility
of both prograde and retrograde precession. There are also non-steadily
precessing solutions with time-varying tilt.
We explore these in Section~\ref{numer} after
discussing the numerical solution method to the basic equation, which is
derived in Section~\ref{eqmot}. Then we apply the results to some of the
known long periods in X-ray binaries and to some similar systems that do
not have observed long periods (Section~\ref{appli}). Finally, we discuss
some implications and limitations of our findings (Section~\ref{discu}) and
summarise our conclusions (Section~\ref{conclu}).

%%%%%%%%%%%%%%%%%%%%%%%%%%%%%%%%%%%%%%%%%%%%%%%%%%%%%%%%%%%%%%%%%%%%%%%%%%%%%%%%
   \section{Equation of motion for an irradiated disc in a binary}
   \label{eqmot}
%%%%%%%%%%%%%%%%%%%%%%%%%%%%%%%%%%%%%%%%%%%%%%%%%%%%%%%%%%%%%%%%%%%%%%%%%%%%%%%%

The accretion disc is assumed to be thin and Keplerian and is divided up into
annuli that interact with each other via viscous forces, which has the
advantage that only the evolution on the longer viscous time scale needs to
be followed. The relevant equation of motion reads
\begin{eqnarray}
   \label{eq:motion}
   \pder{\bL}{t} & = & 
      \frac{3}{R}\pder{}{R}\left[
	 \frac{R^{1/2}}{\Sigma}\pder{}{R}\left(\nu_1\Sigma R^{1/2}\right)\bL
			   \right] \nonumber \\
                &   &
      +\frac{1}{R}\pder{}{R}\left[\left(
         \nu_2R^2\left|\pder{\bl}{R}\right|^2 - \frac{3}{2}\nu_1
	                          \right)\bL
			   \right] \nonumber \\
                &   &
      +\frac{1}{R}\pder{}{R}\left(
         \frac{1}{2}\nu_2R\left|\bL\right|\pder{\bl}{R}
	                    \right) \nonumber \\
		&   &
      +\frac{1}{2\pi R}\tder{\bmath{G}}{R} + \bmath\Omega_{\bf p}\times\bL.
\end{eqnarray}
The first three terms on the righthand side were derived by Papaloizou \&
Pringle (1983, eq.~2.4)\nocite{pp:83} and further discussed by Pringle
(1992, eq.~2.8)\nocite{pring:92}. They conserve angular momentum exactly.
The independent variables are $(R,t)$ and we work in a non-rotating frame
centred on the accreting point mass $M$ (with the $z$ axis normal to the
orbital plane).  $\bl$ is the unit vector
$(\cos\gamma\sin\beta,\sin\gamma\sin\beta,\cos\beta)$ normal to the disc
at radius $R$, so $\beta(R,t)$ is the local tilt angle of the disc
plane and $\gamma(R,t)$ the azimuth of the tilt. $\Sigma$ is the disc
surface density, and $\bL$ is the angular momentum per unit area.  The
assumption of Keplerian rotation implies $\bL=\Sigma\bl\sqrt{GMR}$.
$\nu_1$ is the viscosity associated with the $(R,\phi)$ shear, whereas 
the viscosity associated with the $(R,z)$ shear that damps
the misalignment between neighbouring annuli is $\nu_2$. The term
involving $\bmath{G}$ is the effect of the torque induced by irradiation
from the central object:
\begin{equation}
   \label{eq:radtorq}
   \tder{\bmath{G}}{R}  = 
   \frac{L_\star}{6\pi cR}\frac{\bmath{g_\phi}}{2\pi},
\end{equation}
where the dimensionless vector $\bmath{g_\phi}$ is an integral along the
annulus that describes the purely geometric part of the torque (see
Pringle 1996, 1997)\nocite{pring:96,pring:97}.  $L_\star$ is the
luminosity of the central point source, which is assumed to radiate
isotropically, and each element of the disc that is illuminated by the
central source is assumed to re-radiate the incident radiation
isotropically, causing a pressure perpendicular to the local disc surface.

The last term in \Eq{eq:motion} is the forced external 
precession due to the companion star tide. Let a companion of mass 
$M\sub{c}$ be located in the $XY$ plane at $\bmath{a}=(a\cos\psi,a\sin\psi,0)$.
To lowest order in $R/a$ the torque on an annulus of width d$R$ can be
obtained by integrating the force along an unperturbed circular orbit.
The force on an element of the annulus at position $\bmath{x}$ is
\be{eq:dFc}
   {\rm d}\bmath{F}_{\bf c} = 
   \frac{GM\sub{c}(\bmath{a}-\bmath{x})}{|\bmath{a}-\bmath{x}|^3}
   \Sigma R{\rm d}R{\rm d}\phi.
\ee
The torque on the annulus is then simply obtained by integrating the 
elemental torque contributions along it. Here we expand the force in
powers of $R/a$ and only retain the lowest non-vanishing order of the 
torque:
\begin{equation}
   \label{eq:tidetorq}
   \oint \bmath{x}\times{\rm d}\bmath{F}_{\bf c} =
      \frac{3\pi}{2} GM\sub{c}\Sigma{\rm d}R\frac{R^3}{a^3}
      \left(\begin{array}{c}
	   \sin2\beta\sin\psi\cos(\gamma-\psi) \\
	  \makebox[0in]{$-$\,}\sin2\beta\cos\psi\cos(\gamma-\psi) \\
	   \sin^2\beta\sin2(\gamma-\psi)
	    \end{array}\right)
\end{equation}
Note that the only force taken into account is the gravity of the
companion.  Since our frame is centred on $M$ (but not rotating)
and thus revolves around the centre of mass of the binary, there is
a centrifugal force as well. It is not hard to show that it
contributes no net torque on an annulus. Comparison with the
expression for $\bL$ leads to the expression for the precession frequency:
\begin{eqnarray}
   \bmath{\Omega}_{\rm p} & \equiv & \Omega_{\rm p} \bmath{e}\sub{p} \nonumber\\
   \label{eq:omprec}
   \Omega\sub{p} & = & -\frac{3}{4}\frac{GM\sub{c}}{\Omega a^3} \\
   \bmath{e}\sub{p}& = & \sin\beta\cos(\gamma-\psi)
      \left(\begin{array}{c} \cos\psi \\ \sin\psi \\ 0 \end{array}\right).
      \nonumber
\end{eqnarray}
In this paper, we shall make the approximation that the orbital period
of the binary is small enough compared to the precession period that we
may average the precession over the binary orbital period (i.e.\ over
the angle $\psi$). Averaging and cross-product are not interchangeable, of
course, so we have to average the torque (\Eq{eq:tidetorq}) and then find
a new expression for $\bmath{\Omega}_{\rm p}$. The result is:
\begin{equation}
   \label{eq:omprecav}
   \bmath{\Omega}\sub{p}\spr{av} = \Omega\sub{p}\bmath{e}\sub{p}\spr{av}
    = \Omega\sub{p}\cos\beta\bmath{e}_z
\end{equation}
(See also Katz et~al\ 1982, Papaloizou \& Pringle
1983)\nocite{kamg:82,pp:83}. Note that the factor $\cos\beta$ is not
present in studies that assume the inclination to be small.

For numerical studies we cast the equation in
dimensionless form. The dimensionless variables are $r=R/R_0$,
$\sigma=\Sigma/\Sigma_0$ and $\tau=t/t_0$, with $R_0$, $t_0$ and $\Sigma_0$
values to be determined later. Then we can also set $\bcL = \bL/L_0 =
\bL/\Sigma_0\sqrt{GMR_0}$, $\nu_1 = n_1\nu_{10}$, and $\eta=\nu_2/\nu_1$.
Then if we measure time in units of the viscous time scale
in the disc, i.e.\ $t_0 = R_0^2/\nu_{10}$, we obtain:
\begin{eqnarray}
   \label{eq:motdimless}
   \pder{\bcL}{\tau} & = & 
      \frac{3}{r}\pder{}{r}\left[
	 \frac{r^{1/2}}{\sigma}\pder{}{r}\left(n_1\sigma r^{1/2}\right)\bcL
			   \right] \nonumber \\
                &   &
      +\frac{1}{r}\pder{}{r}\left[n_1\left(
         \eta r^2\left|\pder{\bl}{r}\right|^2 - \frac{3}{2}
	                          \right)\bcL
			   \right] \nonumber \\
                &   &
      +\frac{1}{r}\pder{}{r}\left(
         \frac{1}{2}\eta n_1r\left|\bcL\right|\pder{\bl}{r}
	                    \right) \nonumber \\
		&   &
      +F_\star\frac{\bmath{g_\phi}}{2\pi r} 
      +\omega\sub{p}\bmath{e}\sub{p}\spr{av}\times\bcL.
\end{eqnarray}
where $\omega\sub{p}=\Omega\sub{p}t_0$, and
the dimensionless strength $F_\star$ of the radiation field is given by
\begin{equation}
   \label{eq:fstar}
   F_\star = 
     \frac{L_\star}{6\pi c\Sigma_0R_0^3\Omega_0}\frac{R_0^2}{\nu_{10}}.
\end{equation}

%-------------------------------------------------------------------------------
      \subsection{Analytic considerations}
      \label{resul.anal}
%-------------------------------------------------------------------------------

The stability to warping of a disc described by \Eq{eq:motion} can be
analysed by linearising the disc tilt evolution (eq.~3.5 of Pringle 
1996)\nocite{pring:96} in $\beta$.
Then $\bl=(\beta\cos\gamma,\beta\sin\gamma,1)$ and we can write its
evolution as an equation for $W\equiv\beta{\rm e}^{i\gamma}$.
We further assume that we are far from the disc edges, so that
$\dot{M}\propto \nu_1\Sigma=$ constant and $V_R=\nu_1\Omega^\prime/\Omega$.
Then with $\Sigma\propto \nu_1^{-1}\propto\nu_2^{-1}$
we can write the following equation for $W$:
\be{eq:W}
  \pd{W}{t}=\frac{1}{2}\nu_2\frac{\partial^2W}{\partial R^2} +
            \left(\frac{3\nu_2}{4R}-i\Gamma\right)\pd{W}{R} +i\Omega\sub{p}W,
\ee
where $\Gamma=L_\star/12\pi c\Sigma R^2\Omega$.
The local stability then follows by looking for solutions of the form
$W=W_0\exp i(\sigma t+kR)$, with $kR\gg1$. Substituting in \Eq{eq:W} this
implies
\be{eq:stab}
  \sigma = i[-\Gamma k + \frac{1}{2}\nu_2k^2] + \Omega\sub{p} + 
           \frac{3\nu_2}{4R}.
\ee
The complex part of $\sigma$ is identical to that of Pringle, so neither
the addition of forced precession nor retention of the gradient terms
affects the local linear stability of the disc. Thus, as before, growing modes
occur only for $\Gamma>0$ and $0<k<2\Gamma/\nu_2$ and their line of nodes
follows a leading spiral. At finite amplitudes the  effect of forced 
precession is strong because $\Omega\sub{p}\propto R^{3/2}$, so it will
make the disc precess more rapidly towards the outside and thus
tends to destroy the shape of the growing mode.

%%%%%%%%%%%%%%%%%%%%%%%%%%%%%%%%%%%%%%%%%%%%%%%%%%%%%%%%%%%%%%%%%%%%%%%%%%%%%%%%
   \section{Numerical experiments}
   \label{numer}
%%%%%%%%%%%%%%%%%%%%%%%%%%%%%%%%%%%%%%%%%%%%%%%%%%%%%%%%%%%%%%%%%%%%%%%%%%%%%%%%

%-------------------------------------------------------------------------------
      \subsection{Numerical Method}
      \label{numer.method}
%-------------------------------------------------------------------------------
	The numerical method employed is that described in Pringle
(1997)\nocite{pring:97}, with changes made in order to model accretion
discs in binary systems rather than in AGN. The most important changes
which need to be made are with regard to tidal truncation of the disc at
the outer edge, with regard to the input of mass at some radius other than
the outer edge, and with regard to the tidally forced precession of the
outer disc elements.

	We work initially with 40 grid points between
$r\sub{in} = 1$ and $r\sub{out} = 40$, and use an equally spaced rather
than a logarithmic grid (Section~\ref{numer.res}).  This was chosen so
that the outer edge of the grid would be properly resolved, but has the
disadvantage that the inner disc regions are not well resolved. However,
because of this, the computation time required for each run is relatively
short, and thus we were able to explore a large volume of parameter
space. We report below (Section~\ref{numer.big}) on a few runs carried
out with a more extensive grid. The number of azimuthal zones was normally
68. Increasing this value to 120, which we did in a few key runs, had
no effect on the results. 

	We achieve a zero torque inner boundary condition in the
manner described in Pringle (1997) by removing mass over three zones
centred on $r = 3$. We add material at a constant rate over 3 zones
centred on radius $r\sub{add}$. The material is added to the disc with
specific angular momentum vector corresponding to the material being
added in the orbital plane, and magnitude appropriate for the radius
$r\sub{add}$.

	In order to take account of the tidal truncation of the disc
at radius $r\sub{tide}$ ($< r\sub{out}$), we need to set up a boundary
condition which removes angular momentum at that radius but does not
remove mass. Since the main dependent variable employed is the angular
momentum density $\bcL$, this must of necessity employ some
approximation. We find that the simplest method of achieving the
desired effect is to calculate at each time step the amount of mass
external to $r\sub{tide}$, to set the contents of those zones to zero,
and to add the mass to the zone just internal to $r\sub{tide}$, with
the same specific angular momentum as the matter already in that
zone. In this way it was intended to minimise the chance of the outer
tidal truncation resulting in any (unphysical) instability. Indeed no
such instabilities were observed. This boundary condition should
correspond to the combined conditions of $v_R = 0$ and
$\partial\bl/\partial R = 0$ at $r = r\sub{tide}$.

	The forced precession caused by the tidal potential field is
handled as follows. Each annulus is precessed at each
time step about the vector (0,0,1) at a rate $\omega\sub{p} =
\omega\sub{p0} r^{3/2}$.  Thus we have ignored the factor of
$\cos\beta$ present in the forced precession rate (Equation 1.6). This
could be simply included, but should be negligible for the kind of
solutions we are looking for. The time step is limited to ensure that
any annulus is not precessed by more than $2 \pi /10$ in any one
time step. In practice the time step considerations for stability of the
numerical method ensure that the precession rate is not the dominant
factor in limiting the time step.

	We assume a viscosity of the form:
\be{eq:numvisc}
	\nu_1 = \nu_{10} r^{3/4},
\ee
which is chosen to approximate the viscosity dependence in such discs
(see eq.~\ref{eq:visc}). We take $\eta = 1$. The disc is set up initially with a
surface density appropriate to a steady disc with this viscosity and
with the appropriate boundary conditions (viz.\ $\Sigma=0$ at $r=1$, $v_R=0$
at $r=r\sub{tide}$, and a dimensionless mass input rate of unity
at $r=r\sub{add}$). The
disc is given an initial warp of the form of a prograde spiral which
is such that $\beta$ increases from zero at $r=1$ to $\beta=0.1$ at $r=40$,
with $\gamma$ increasing by an angle of one radian over the same
distance.

%-------------------------------------------------------------------------------
      \subsection{Numerical Results}
      \label{numer.res}
%-------------------------------------------------------------------------------

	We have investigated the non-linear behaviour of irradiated
discs in binary systems, taking shadowing fully into account. Our aim
has been to discover what regions of parameter space (if any) give
rise to discs with a finite, and (reasonably) steady tilt angle of the
kind which might be relevant to explanations of systems like Her
X-1/HZ Her. Thus our initial aim was to search for the kind of
`mode-like' behaviour discussed by Maloney \& Begelman (1997)\nocite{mb:97}.
We found, however,
that the time-dependence of binary accretion discs gives rise to a
richer and more varied behaviour than can be described by simple `modes'.

%-------------------------------------------------------------------------------
      \subsubsection{$r\sub{add} = 10, r\sub{tide} = 30, 
                   \omega\sub{p0} = 0$}
      \label{numer.res.1}
%-------------------------------------------------------------------------------

\begin{figure}
    \epsfxsize=\columnwidth\epsfbox{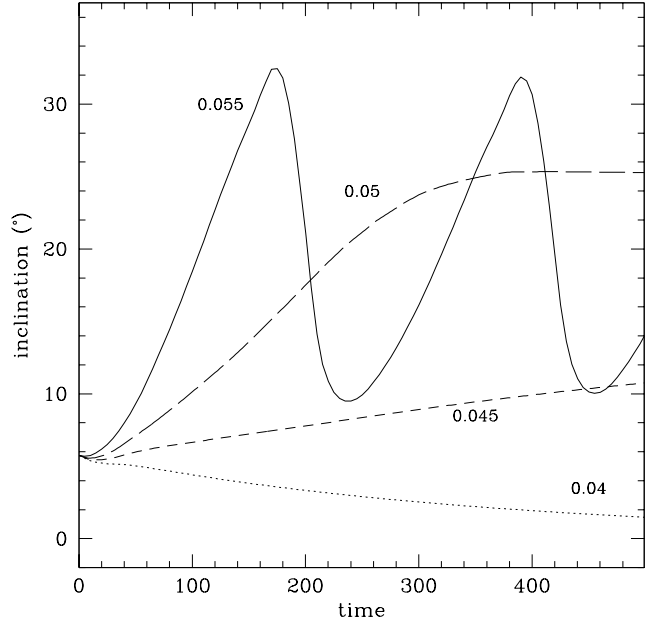}
    \caption{The behaviour of the disc inclination at $r=26$, just inside
	     $r\sub{tide}$, for
	     different values of $F_\star$ and using the small grid
	     (Section~\ref{numer.res}). With increasing $F_\star$, the
	     disc first is stable, then grows to a finite and constant tilt
	     that is greater for higher values of $F_\star$.
	     The oscillations at still higher $F_\star$ are an artefact
	     of the poorly resolved inner disc.
	     \label{fi:tiltfstar}
	    }
\end{figure}

	With an eye on the parameters of the binary systems we are
interested in as shown in Table~\ref{ta:num}, we choose 
$r\sub{add}/r\sub{tide}= 0.33$. We
initially set the forced precession rate to zero, and investigate what
values of the disc luminosity ($F_\star$) give rise to relevant
behaviour. In Figure~\ref{fi:tiltfstar}
we show the behaviour of the inclination of
the outer disc edge as a function of dimensionless time for various
values of $F_\star$. We find that for $F_\star\le 0.04$
the disc irradiation is not
strong enough to lead to instability and the disc flattens into the
orbital plane.

	For $F_\star=0.045$ the disc tilt settles down to a steady functional
form, with $\beta$ small at the inside, and $\beta^\prime$ positive at most
radii, reaching a value of $\beta=0.275$ (corresponding to an angle of
$16^\circ$) at the outside (i.e.\ at $r=r\sub{tide}$). The disc is warped in the
shape of a prograde spiral, as is to be expected for a disc unstable
to self-irradiation (Pringle 1996).  At the same time the whole disc
precesses steadily in a retrograde direction with a dimensionless
period of $P\sub{p}= 58$. The direction of precession is retrograde because
of the shape of the disc. The sign of $\beta^\prime$ determines which face of
the disc is illuminated by the central source. A positive $\beta^\prime$
leads to retrograde precession (Pringle 1996). We note that this
contrasts with the AGN discs discussed in Pringle 1997. There the
outer boundary condition ensured that the outer edge of the disc
remained in the initial disc plane. Then if the inner disc is tilted,
much of the disc is likely to have negative $\beta^\prime$, and therefore
precesses in a prograde fashion. In the binary case the edge of the disc is
free in the sense that no torque acts there to change the tilt of the
disc.

	Similar behaviour occurs when $F_\star=0.05$, although in this case
$\beta$ at the outside settles down to the larger value of $\beta=0.44$
(corresponding to an angle of $25^\circ$) and to a more rapid retrograde
precession rate with a period $P\sub{p}=48$. 

	When the effect of radiation is further increased,
the disc behaviour is no longer steady. The behaviour of $\beta$ at the
outer radius is shown in Figure~\ref{fi:tiltfstar}.
The behaviour is in the form of a limit cycle.  We note here
that for simplicity we have assumed that $F_\star$ is constant
throughout each run. However, in reality, since the radiation 
illuminating the disc comes directly from the accretion rate at the
disc centre, $F_\star$ is likely to vary with time when the disc
displays such non-steady behaviour. Thus the actual disc behaviour
is likely to be yet more complicated than we find here.

%-------------------------------------------------------------------------------
      \subsubsection{$r\sub{add} = 10, r\sub{tide} = 30, 
                   F_\star = 0.05$}
      \label{numer.res.2}
%-------------------------------------------------------------------------------

	In this section we describe the effect on the steady
precessing solutions of adding retrograde forced precession of the
form induced by tidal torques from a companion. The magnitude of the
forced precession is described by the parameter $\omega\sub{p0}$ 
(eq.~\ref{eq:omegap}).
With zero forced precession, $\omega\sub{p0}=0$, we have seen 
(Section~\ref{numer.res.1})
that the disc settles down in the shape of a prograde spiral,
with $\beta$ an increasing function of $r$, reaching $\beta=0.44$ at the
outer edge, and precessing steadily in a retrograde direction with
period $P\sub{p} = 48$.

	When a small amount of retrograde forced precession is added,
the disc settles down to a similar behaviour as for $\omega\sub{p0}=0$, but
with the final value of $\beta(r\sub{tide})$ slightly increased, and the
precession rate also increased. Thus we find that for
$\omega\sub{p0}=-0.0002$, $\beta(r\sub{tide})=0.54$ and $P\sub{p}=40$; for
$\omega\sub{p0}=-0.0005$, $\beta(r\sub{tide})=0.63$ and $P\sub{p}=35$, and for
$\omega\sub{p0}=-0.001$, $\beta(r\sub{tide})=0.56$ and $P\sub{p}=29$. 
However, as the
forced precession rate is increased further the disc instability is
removed. Thus, for $\omega\sub{p0}=-0.002$, we find that $\beta(r\sub{tide})$
tends
to zero, and the disc settles down into the orbital plane. We suggest
that this behaviour comes about because in order to be unstable the
disc must take the form of a prograde spiral (Pringle 1996). However
strong retrograde forced precession, which is differential in the
sense that the precession rate increases with radius (here $\propto
r^{3/2}$), tends to unwind the prograde spiral and
thus acts to prevent the instability from occurring.

%-------------------------------------------------------------------------------
      \subsubsection{$r\sub{add} = 10, r\sub{tide} = 30, 
                   F_\star = 0.09$}
      \label{numer.res.3}
%-------------------------------------------------------------------------------

	Here we investigate the effect of adding forced retrograde
precession on the solutions discovered in Section~\ref{numer.res.1}
 which take the form of limit cycles. We therefore investigated solutions
with $F_\star=0.09$, and with $\omega\sub{p0}=-0.001$, $-0.002$, $-0.003$
and $-0.004$.  Increasing the forced precession rate seems to have little
effect on the amplitude of the limit cycle, until at the value of $-0.004$
the instability is quenched altogether, and the disc settles into the
orbital plane. However the period of the limit cycle is affected, and
decreases with increasing $\omega\sub{p0}$, from a value of 200 for
$\omega\sub{p0}=0$ to a value of about 120 for $\omega\sub{p0}=-0.003$.
While these limit-cycle solutions are partly an artefact of the small grid,
the damping of even these high-luminosity solutions illustrates the
how powerful forced precession is in suppressing disc tilts.

%-------------------------------------------------------------------------------
      \subsubsection{$r\sub{add} = 20, r\sub{tide} = 30, 
                   \omega\sub{p0}=0$}
      \label{numer.res.4}
%-------------------------------------------------------------------------------

	One effect of adding material at $r\sub{add}=10$ in the
previous sections was to help pin the disc towards the orbital plane
at that radius. This helped to control the behaviour of the disc
within that radius (see also Section 3.3), and to allow the outer
regions of the disc (a factor of three in radius) to evolve freely
with regard to tilt. Since the self-irradiation warping instability
acts more strongly at larger radii, it was always the outer regions of
the disc which responded most to the radiation flux from the central
object. In this section we describe the effect of changing the radius
at which mass is added to the disc from $r\sub{add}=10$ to
$r\sub{add}=20$.  This has two major effects. First, adding matter
closer to the outside has the effect of tending to pin the disc into
the orbital plane at the outside, while allowing the inner disc
regions to tilt (see also Section~\ref{numer.big}).
Second, a steady disc inside the radius $r\sub{add}$
has $\nu \Sigma$ constant at radii inside $r\sub{add}$, whereas a
steady disc with a $v_R=0$ outer boundary condition has $\nu \Sigma
\propto r^{-1/2}$ for radii outside $r\sub{add}$.  Thus for a given
accretion rate, since $\nu$ is typically an increasing function of
both $\Sigma$ and $r$, the outer parts of the disc become less massive
as $r\sub{add}$ is decreased. Thus by increasing $r\sub{add}$ from 10
to 20, we expect the radiation instability to require larger values of
$F_\star$.

\begin{figure}
    \epsfxsize=\columnwidth\epsfbox{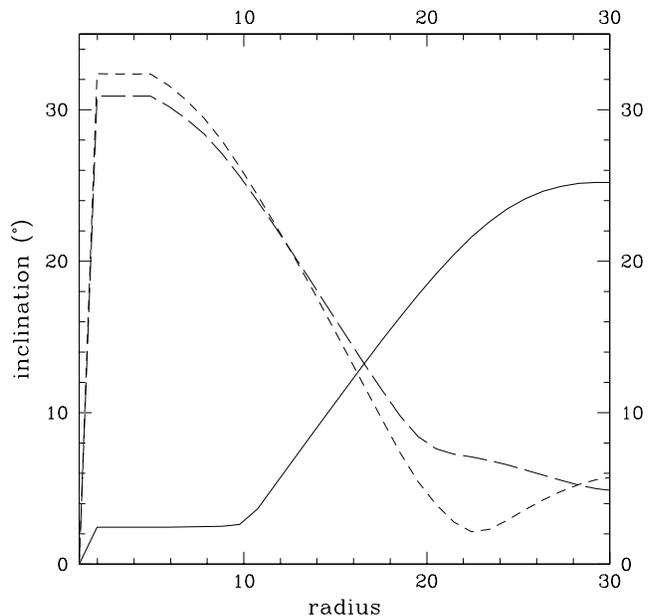}
    \caption{Comparison of the variation of the disc tilt $\beta$
             with radius for mass addition well inside the outer
             radius ($r\sub{add}=10$, solid curve) and mass addition
             nearer the outer edge ($r\sub{add}=20$, dashed curves).
             The former, having $\beta^\prime>0$ in most of the disc, precesses
	     retrogradely, whereas the latter precesses progradely.
	     Note that when $r\sub{add}=20$ the disc 
	     is not a fixed-shape object precessing at a single rate (see text).
	     The two curves span approximately the range of shapes it displays.
	     \label{fi:betofr}
	    }
\end{figure}

	For values of $F_\star\le 0.15$, we find that the disc is stable to
self-irradiation. For values of $F_\star\ge 0.25$ the inner parts of the
disc turn over completely in the manner discussed for the AGN discs in
Pringle 1997. For $F_\star=0.2$ the disc does tend to a fairly steady
configuration. However the shape of the disc now differs from those
discussed above in the sense that the disc tilt $\beta$ is largest at
the inside, and decreases with radius (i.e.\ $\beta^\prime$ is predominantly
negative; Fig.~\ref{fi:betofr}). 
Moreover the inner and outer parts of the disc behave
in quite different and independent manners. The inclination at all
radii in the disc oscillates more or less in phase with a period of
about $P\sub{inc}=10$. The innermost radii oscillate between $\beta=0.53$ and
0.57, and the outermost radii oscillate between $\beta=0.06$ and
0.12. At radius $r=24$, the oscillation in $\beta$ varies between 0 and
0.14, and it is this radius which appears to separate the inner and
outer parts of the disc. The inner part of the disc (roughly those
radii $r<24$) has an inclination which decreases with radius, and which
precesses in a prograde direction with a period of about $P\sub{in}=12$. The
outer parts of the disc (roughly radii in the range $24<r<30$) have (on
a time average) an inclination $\beta$ which increases with radius and
precess in a retrograde direction with a period of about $P\sub{out}=50$. The
precession in the inner regions proceeds at a more or less steady
rate. However the azimuth of the tilt of the outer edge of the disc
stays almost constant for each oscillation period of the inclination,
and then jumps by a certain amount (which then gives the mean outer
precession period). It should be noted that the oscillation frequency
of the inclination is the sum of the moduli of the precession
frequencies of the inner and outer parts of the disc (i.e.\ $P\sub{inc}^{-1} =
P\sub{in}^{-1} + P\sub{out}^{-1}$).

%-------------------------------------------------------------------------------
      \subsubsection{$r\sub{add} = 20, r\sub{tide} = 30, 
                   F_\star=0.2$}
      \label{numer.res.5}
%-------------------------------------------------------------------------------

	We now investigate the effect of adding retrograde forced
precession to the $F_\star=0.2$ disc which was described in the previous
section. As the magnitude of $\omega\sub{p0}$ increases from zero, the
behaviour of the disc stays initially the same, except that the
inclination decreases, the inner (prograde) precession period,
$P\sub{in}$, increases, and the outer precession period, $P\sub{out}$,
decreases. Thus, in comparison with $\beta(r\sub{in})=0.55$,
$P\sub{in}=12$ and $P\sub{out}=50$ for $\omega\sub{p0}=0$, we find that
for $\omega\sub{p0}=-0.0015$, $\beta(r\sub{in})=0.46$, $P\sub{in}=14$,
$P\sub{out}=20$; for $\omega\sub{p0}= -0.004$, $\beta(r\sub{in})=0.32$,
$P\sub{in}=20$, $P\sub{out}=11$; for $\omega\sub{p0}=-0.006$,
$\beta(r\sub{in})=0.2$, $P\sub{in}=26$, $P\sub{out}=6$; and for
$\omega\sub{p0}=-0.007$, $\beta(r\sub{in})=0.16$, $P\sub{in}=40$ and
$P\sub{out}=4.5$. In addition, as the size of $\omega\sub{p0}$ is
increased, the radius which separates the inner disc and outer disc
behaviours (the radius at which the $\beta$ oscillation passes through
zero) moves outwards until at $\omega\sub{p0}=-0.006$ it is at the outer
edge to within the resolution of the grid.  However when $\omega\sub{p0}$
is increased further to $-0.008$ the whole disc now settles to a constant
$\beta$ profile, which is positive at all radii, and has
$\beta(r\sub{in})=0.12$, and $\beta$ decreasing with radius. The disc as a
whole precesses in a prograde direction with period $P\sub{p}=22$. The
shape of the disc is such that it has a prograde spiral inside
$r\sub{add}$, and a retrograde spiral outside $r\sub{add}$.

%------------------------------------------------------------------------------
	\subsection{Numerical results on a more extensive grid}
	\label{numer.big}
%------------------------------------------------------------------------------

	Since all the numerical results above were computed on a grid
	which had limited resolution in the inner regions, we felt
	that it was necessary to explore the limitations of such a
	procedure. We use a logarithmic grid consisting of 80 grid
	points between $R\sub{in} = 0.136$ and $R\sub{tide} = 30$. We still add
	material over 3 grid points centred on $R\sub{add} = 10$, and we
	remove material over 3 grid points centred on $R = 0.155$. In
	this manner the outer part of the grid between $R\sub{add}$ and
	$R\sub{tide}$ is reasonably well resolved (15 grid points), and the
	inner disc region extends inward of $R\sub{add}$ by almost two
	orders of magnitude, so that the behaviour of the inner parts
	of the grid can now be modelled more accurately.

	Because the grid is now more extensive, and especially
	because the grid now extends to smaller radii (where viscous
	time scales are shorter) the computational run times are now
	longer by about two orders of magnitude. Thus we have limited
	ourselves to a few representative examples for comparison with
	the results in the previous section. We find that the
	instability sets in at values of $F_\star$ which are larger by
	about a factor of two. This comes about because in a steady
	accretion disc with an inner boundary condition corresponding
	to vanishing surface density (i.e. zero torque), it is the
	quantity $\nu \Sigma [1 - (r/r\sub{in})^{1/2}]$ which is constant
	with radius, rather than just $\nu \Sigma$. Thus the effect of
	moving the inner boundary inwards is to increase the surface
	density of the outer disc by about a factor of two.

\begin{figure*}
    \epsfxsize=\textwidth\epsfbox{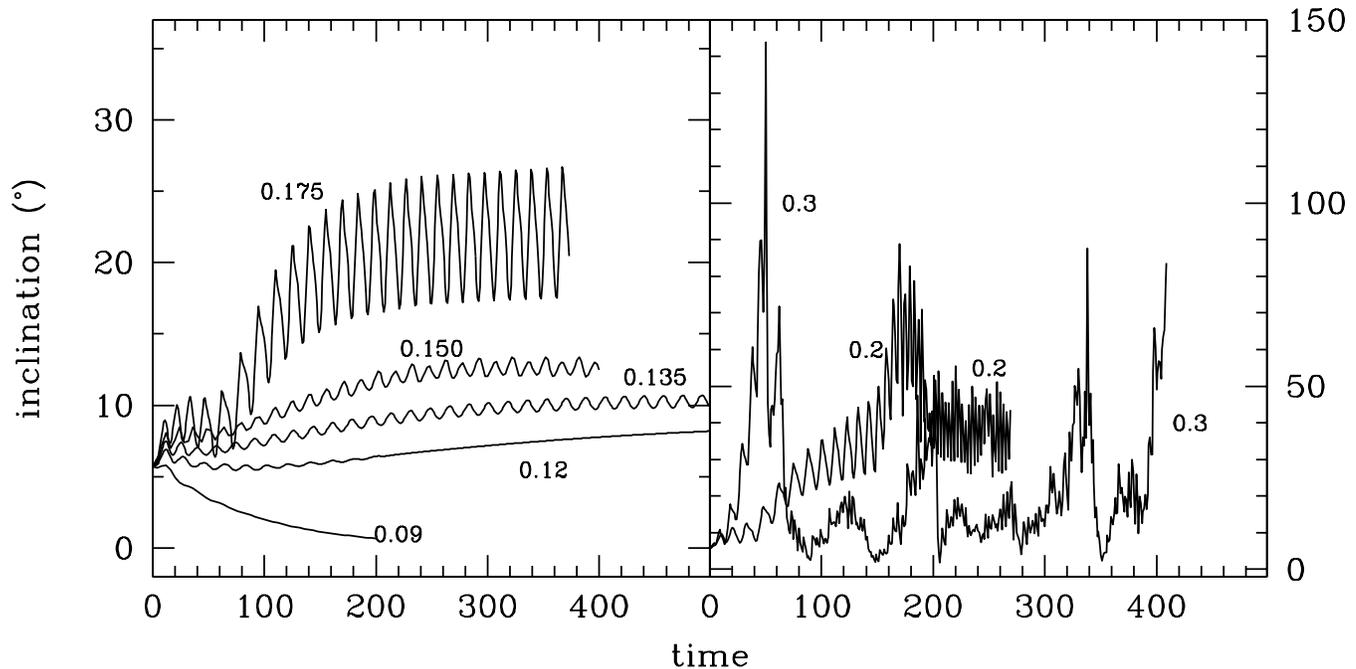} 
    \caption{The behaviour of the disc inclination at $r=26$, just inside
	     $r\sub{tide}$, for different values of $F_\star$. With
	     increasing $F_\star$, the disc first is stable, then grows to
	     a finite and constant tilt that is greater for higher values
	     of $F_\star$. The oscillations around the stable level are
	     due to the fact that the inner and outer disc precess in
	     opposite directions once the unstable region in the disc
	     includes enough of the disc inside $r\sub{add}$, so that the
	     solution is no longer stationary. At even greater $F_\star$
	     the disc tilts through 90 degrees and the behaviour becomes
	     chaotic. This is the regime previously discussed by Pringle
	     (1997) for AGN. 
	     \label{fi:bigtiltfstar}
	     }
\end{figure*}

\begin{figure*}
    \epsfxsize=\textwidth\epsfbox{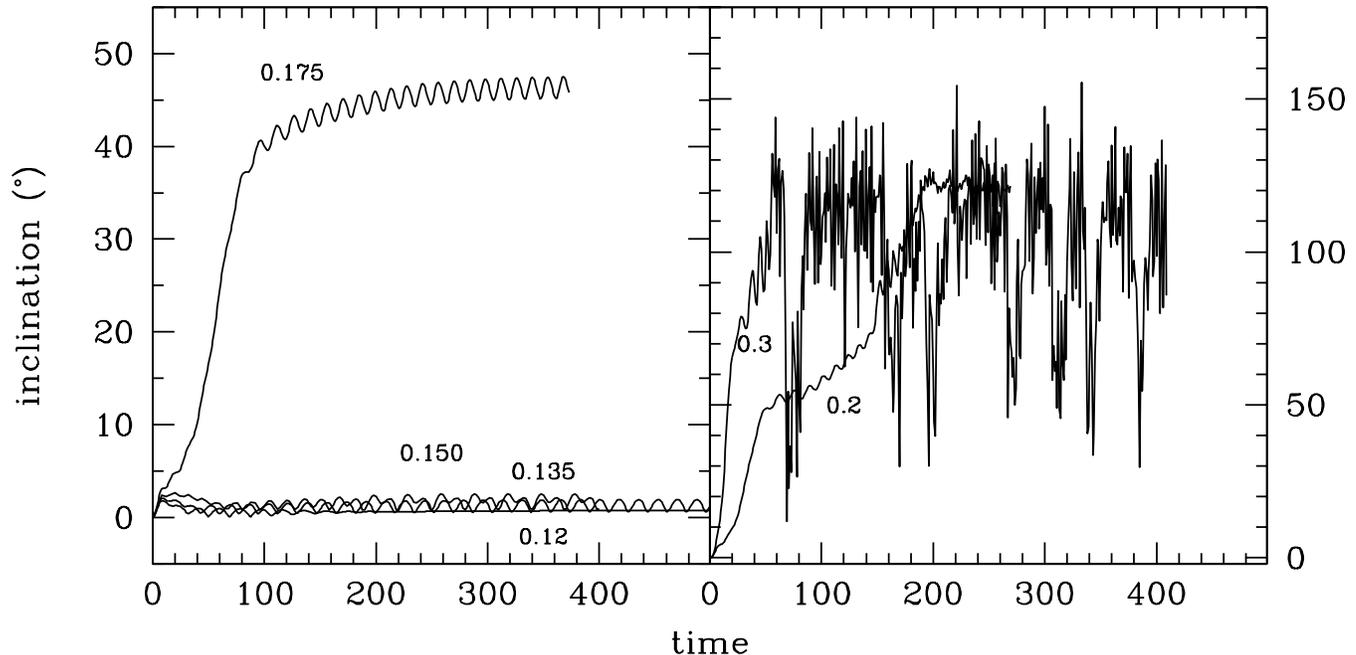}
    \caption{The behaviour of the disc inclination at $r=2.5$, well inside
	     $r\sub{add}$, for different values of $F_\star$.  Note the
	     larger tilts in comparison with the outer disc and the fact
	     that at large $F_\star$ the inner disc is mostly
	     counter-rotating.
	     \label{fi:bigtiltfstarin}
	     }
\end{figure*}

	Since the tidally induced precession does not seem to play a
	strong role for those discs which are relevant to the X-ray
	binaries we are interested in here, we have taken $\omega\sub{p}
	= 0$. In Figure~\ref{fi:bigtiltfstar}
	we show the behaviour of the disc
	inclination at $r=26$ for $F_\star$ = 0.09, 0.12, 0.135, 0.15, 0.175,
	0.2 and
	0.3. As can be seen the disc with $F_\star = 0.09$ is
	stable. The disc with $F_\star = 0.12$ eventually precesses in
	a steady fashion in a retrograde direction with a period of
	about 35, and the disc inclination at the outer edge
	settles down to a value of 0.15 (i.e. about 8.5 degrees). The
	disc with $F_\star = 0.15$ settles down to a solution in which
	the inclination and oscillates
	with a period of about 15 and semi-amplitude 5 per cent, about a
	steady value for the inclination of 0.25 (14 degrees). The precession
	period is 30 for the outer disc. The inner disc has inclination
	near zero, with small oscillations of the same period as the outer
	disc (Figure~\ref{fi:bigtiltfstarin}). The period of these oscillations
	is the beat period between the inner and outer disc periods, as with
	the small grid. Their cause is simply that the region where the
	outer and inner disc join (just outside $r\sub{add}$) tries to
	adjust to the tilts of both sides. This it cannot do
	simultaneously, of course, and the tilt of this zone oscillates
	between nearly zero and a finite value, depending on the relative
	phase between the inner and outer disc solutions. This oscillation
	is then communicated throughout the disc, with an amplitude that
	decreases away from the contact zone.

	For greater $F_\star$, the inner disc suddenly gets a substantial
	tilt as well, presumably because it too is now unstable (the
	sudden transition is caused by the fact that the inner radius of
	the unstable region decreases as the inverse square of $F_\star$,
	see Pringle 1997).  The discs with $F_\star = 0.2$ and 0.3 vary
	chaotically, with the tilt of the inner disc usually greater than
	90 degrees, and that of the outer disc less.  The outer disc still
	precesses retrogradely on average, with roughly the same period as
	before (25--35), but with irregularities superimposed. The azimuth
	of the inner disc tilt wanders irregularly without any long-term
	trend. 

	In conclusion we find that the results obtained with the
	larger grid are fully consistent with the behaviour found with
	the cruder grid. Thus the results presented in Section 3.2 are
	expected to be reasonably representative. The main differences
	appear to be that the larger grid has a larger surface density
	(for the reasons explained above) and so goes unstable for
	somewhat larger values of $F_\star$, that the range of
	$F_\star$ for which the disc displays steady, or nearly
	steady, precessing tilted behaviour is somewhat larger, and
	that the tilt angles reached by the outer disc are somewhat
	smaller. In order to check the grid-dependence of these large
        simulations we further
	doubled the number of radial and azimuthal grid points in a few
	key simulations;
	no significant changes in the results were observed.

%-------------------------------------------------------------------------------
      \subsection{Precession periods and stability}
      \label{numer.pp}
%-------------------------------------------------------------------------------

In order to see how well simple estimates work, we now compare the precession
periods obtained in the simulations with expected values.
From \Eq{eq:motdimless}, one obtains a precession time scale estimate, taking
$g_\phi/2\pi$ to be unity:
\be{eq:tp1}
   t\sub{ \Gamma p} = \frac{{\cal L} r}{F_*},
\ee
where $r = R/R_0$. Now ${\cal L} = \sigma r^{1/2}$, where 
$\sigma=\Sigma/\Sigma_0$, and thus at the outside edge of the disc,
\be{eq:tp2}
  P\sub{p} = 2\pi t\sub{ \Gamma p} = \frac{2\pi}{F_*}\,\sigma r^{3/2}.
\ee
We now have to take the quantities from simulations to compare
the left and righthand sides numerically. Using the results from
\Sect{numer.res.1}, we have, e.g.\ a numerical period of 48 when
$F_*=0.05$. At the outer edge of this simulation, $r=30.25$ and
$\sigma=2.26\times10^{-3}$, giving a predicted period from
\Eq{eq:tp2} of $P\sub{p}=47$, in very good agreement with the value seen
in the simulation. For the run with $F_*=0.045$ we have $\sigma=2.76\times10^{-3}$,
and hence compute $P\sub{p}=64$, compared with the actual simulation
value of 58. It is clear that the radiative time scale at the outer edge
of the grid predicts the numerical precession period quite accurately.
The high precision is doubtless somewhat fortuitous, since shadowing will
tend to lengthen the period at the grid edge somewhat ($g_\phi/2\pi<1$),
whereas the fact that the pattern speed must be an average over a finite
range of radii in the disc tends to shorten it, because the radiative 
time scale becomes shorter at smaller radii.

The critical number for stability is the ratio of radiative to viscous 
time scales at the outer disc edge. In the scaled equations, we have
\be{eq:gamma}
   \gamma = \frac{t_\Gamma}{t_{\nu_2}} = \frac{\sigma}{F_*}\,
            \frac{\nu_{20}}{\nu_{10}}\,r^{1/4}.
\ee
In analogy with Pringle (1997, eq.3.9) we can then define a critical value
$\gamma\sub{crit}$, such that discs with $\gamma<\gamma\sub{crit}$ will be
unstable to radiative warping. Since its precise value will depend on 
boundary conditions, it has to be evaluated from appropriate numerical
simulations. There was a difference in the critical value of $F_*$ between
the large and small grids we explored, and we surmised that this was due
to their different outer-edge surface densities. The evaluation of
$\gamma\sub{crit}$ should be independent of this. For the small grids, the
critical value of $F_*$ is 0.045, and with $\sigma=2.76\times10^{-3}$ at $r=30.25$,
we deduce $\gamma\sub{crit}=0.106$. For the large grids, the smallest 
unstable $F_*$ is 0.09, and for this simulation $\sigma=5.28\times10^{-3}$ at 
$r=29.86$, resulting in $\gamma\sub{crit}=0.102$. The excellent agreement
between the two cases confirms that we understand the difference between the
stability in the large and small-grid simulations. The fact that we find
a three times smaller $\gamma\sub{crit}$ than Pringle (1997) did for
simulations geared to AGN (i.e.\ without a tidal truncation on the disc)
does demonstrate the need for re-determining $\gamma\sub{crit}$ for
different types of system if one wants to go beyond order of magnitude
estimates (see also Maloney, Begelman \& Novak, 1998)\nocite{mbn:98}.

%-------------------------------------------------------------------------------
      \subsection{Summary of numerical results}
      \label{numer.res.sum}
%-------------------------------------------------------------------------------
	Since we have presented the results of a variety of numerical
computations, using two different numerical grids, it would seem
prudent at this stage to provide a brief summary of the basic results
we feel that we have learnt from them. 

	The results from both grids agree that for values of $F_\star$
that are too small the disc is stable against the tilting instability.
They further agree that for values of $F_\star$ above the critical
value for instability by not more than about 30 per cent, the disc
takes on a fixed shape, with inclination increasing to (at most)
around 20 degrees at the outside (dependent on $F_\star$) and in the
form of a prograde spiral, and precesses steadily in a retrograde
direction (at a rate also dependent on $F_\star$). The extent of the outer
part of the disc which is able to display this behaviour depends to some
extend on the radius at which mass is added to the disc. For values of
$F_\star$ slightly larger than this, the disc displays similar behaviour
in the mean (i.e fixed disc shape and steady retrograde precession), but
has an oscillatory behaviour superimposed upon this due to the fact that
the inner and outer disc precess in opposite directions, which means that
in the zone where they meet there is no stationary solution and the tilt
oscillates at the beat period between the inner and outer disc precession
periods. This oscillation is communicated throughout the disc, with an
amplitude that decreases away from the contact region. 

	We have used the smaller grid, which has higher spatial
resolution at the outer disc edge, in order to investigate the effects
of tidally induced differential (retrograde) precession. The basic
effect of differential precession on a tilted disc is to flatten the
disc out into the plane normal to the precession axis (e.g. Pringle,
1992; Scheuer \& Feiler 1996)\nocite{pring:92,sf:96},
and this seems to be the case here.
Note that even the non-linear behaviour which occurs for large values
of $F_\star$ is damped by the imposition of forced differential
precession. 

	The results from the two grids do not agree on the non-linear
development of the disc tilt for large values of $F_\star$. The large
grid, extending a factor of over 200 in radius shows that, as was
found for discs in an AGN type environment (Pringle 1997), the disc
can become completely inverted for large enough values of
$F_\star$. The disc then displays chaotic behaviour as a result of the
interaction between disc self-shadowing, which stabilises the outer
disc, and the viscous time-delay between the outer and inner disc
radii. In contrast the small grid indicates that the disc undergoes
periodic relaxation oscillations. It seems likely that this behaviour
is the result of too few grid points being available to provide an
sufficiently accurate description of the disc shape and dynamics.
The stability analysis and precession periods for marginally unstable
discs, however, agree very well between the large and small grids if
one accounts for their difference in surface density.
	
%%%%%%%%%%%%%%%%%%%%%%%%%%%%%%%%%%%%%%%%%%%%%%%%%%%%%%%%%%%%%%%%%%%%%%%%%%%%%%%%
   \section{Application to some observed systems}
   \label{appli}
%%%%%%%%%%%%%%%%%%%%%%%%%%%%%%%%%%%%%%%%%%%%%%%%%%%%%%%%%%%%%%%%%%%%%%%%%%%%%%%%

%-------------------------------------------------------------------------------
      \subsection{Numerical values of the coefficients}
      \label{appli.numer}
%-------------------------------------------------------------------------------
         \subsubsection{Viscosity}
         \label{appli.numer.visc}
%-------------------------------------------------------------------------------

As usual, the viscosity in an accretion disc is very uncertain. We shall
make the standard assumption of an $\alpha$ disc (Shakura \& Sunyaev
1973)\nocite{ss:73} which radiates the generated heat locally. In the range
of temperatures and densities relevant to the outer regions of accretion
discs in bright X-ray binaries, the dominant radiative opacity is bound-free,
$\kappa=\kappa_0\rho T^{-7/2}$. For solar abundance material,
$\kappa_0=1.5\times10^{24}\un{g}{-2}\un{cm}{5}\un{K}{7/2}$. Using the
standard assumptions we get the viscosity as a function of $R$ and $\Sigma$,
from which we eliminate $\Sigma$ using $\nu\Sigma=\dot{M}/3\pi$ (i.e.\
assuming we are far from the disc edge). This gives
\begin{equation}
   \label{eq:visc}
   \nu = 5.88\times10^{15}\,\,\alpha^{4/5}
         \dot{M}^{3/10}_{-8} M_{1.4}^{-1/4}R_{11}^{3/4}\un{cm}{2}\un{g}{-1}.
\end{equation}
($\dot{M}_{-8}=\dot{M}/10^{-8}\msun\un{yr}{-1}$; $M_{1.4}=M/1.4\msun$;
$R_{11}=R/10^{11}\un{cm}{}$). Since we measure time in units of the
viscous time at the inner edge of the disc only the radial scaling enters
into our calculations directly.

When mass is added at a radius $R\sub{add}$ which is significantly
less than the outer disc radius $R\sub{tide}$ the usual steady-state
disc assumption $3\pi\nu\Sigma=\dot{M}$ does not apply for
$R>R\sub{add}$.  Instead, we have a regime of zero radial velocity,
which implies $\nu\Sigma R^3\Omega^\prime=$const.  (Pringle
1996\nocite{pring:96}; prime denotes radial differentiation).  This
results in $\nu\Sigma\propto R^{-1/2}$. We can then compute the disc
temperature by equating half the local dissipation rate,
$\frac{1}{2}\nu\Sigma(R\Omega^\prime)^2$ to the local emission rate
from each disc surface. Since at its inner edge this part of the disc
does connect to a regime where $\Sigma$ follows from the accretion
rate in the usual way we can still set the normalisation of the
outer-disc surface density from the accretion rate.  The net result is
the same as above (eq.~\ref{eq:visc}), but with an additional factor
$(R/R\sub{add})^{-3/20}$.  We do not consider any discs with
$R\sub{tide}/R\sub{add}>4$, so we have ignored the small correction to
the viscosity in our calculations. Note that the surface density is
significantly affected, changing from $\Sigma\propto R^{-3/4}$ inside
$R\sub{add}$ to $\Sigma\propto R^{-11/10}$ outside, so that the outer
disc becomes significantly lighter and more susceptible to radiative
warping when mass is injected well within $R\sub{tide}$.

%-------------------------------------------------------------------------------
         \subsubsection{Disc size and mass input}
         \label{appli.numer.size}
%-------------------------------------------------------------------------------

In a binary system the accretion stream emerging from the companion at L$_1$
initially has a non-circular orbit, but once it self-intersects it
settles in a circular orbit with radius $R\sub{J}$ which has the same
specific angular momentum about the accreting object as the incoming stream.
This radius is tabulated by Flannery (1975)\nocite{flann:75} for
$0.053<q<19$, where $q\equiv M\sub{donor}/M\sub{accretor}$.
We find that his calculations are well fitted by
\begin{equation}
   \label{eq:rJ}
     r\sub{J} = \frac{R_J}{a} = 0.085\,q^{-0.43}(1+0.47q^2)^{0.1},
\end{equation}
where $a$ is the semi-major axis of the orbit. On a viscous time scale,
the ring spreads inwards and outwards; the outward expansion is halted by
tidal angular-momentum extraction when the disc approaches the Roche
radius. We adopt the outer radii calculated by Papaloizou and Pringle
(1977)\nocite{pp:77} as a function of mass ratio, which to a good
approximation give $R\sub{tide}=0.87R\sub{L}$; here $R\sub{L}$ is the
volume-average Roche radius of the accretor, for which we use the
approximation formula by Eggleton (1983)\nocite{eggle:83}. 
The inner edge of the disc
is either near the surface of the compact object or near the magnetopause,
and it is at least 3 orders of magnitude smaller than $R\sub{tide}$.
Simulations with very large values of $R\sub{tide}/R\sub{in}$ are very
time consuming. Setting $R\sub{tide}/R\sub{in}=30-40$ in the numerical studies
proved adequate, although we undertake some calculations with
$R\sub{tide}/R\sub{in}= 220$ (Section~\ref{numer.big}).

An important issue is where matter enters the disc, and here there is a
clear hysteresis. If the disc is stable, mass enters it at the outer edge
because it is stopped from further infall by the matter already present.
But if the disc is tilted out of the plane, the incoming stream does not
stop until it hits the line of nodes or reaches the point of closest
approach for a ballistic orbit, approximately $R_J/2$. In practice,
therefore, there is a variation of the input radius with the relative
phase between orbit and precession that introduces extra short-term
variations in the disc.  In the spirit of our present approximation
of neglecting these rapid variations, we add matter to the disc in a
small region around $R_J$, where most of the mass would be added in a
more realistic case. Because a flat disc with input at the edge has a
significantly higher surface density than the corresponding tilted disc,
this means that X-ray sources can display a hysteresis effect: if a binary
with a long period temporarily decreases in brightness, for whatever reason,
and remains less luminous for a few outer-disc viscous times, it will not
necessarily return to its long period when the luminosity is restored to
the previous level, because the now higher density means that a higher
luminosity is needed to re-establish the warp than was needed to maintain
the previously established one.

%-------------------------------------------------------------------------------
         \subsubsection{Time scales}
         \label{appli.numer.time}
%-------------------------------------------------------------------------------

There are three time scales that characterise the competing processes of
viscous damping, forced precession, and radiative growth. In the disc
regime $R<R\sub{add}$ and for the above viscosity, their values are
\begin{eqnarray}
   \label{eq:tnu2}
   t_{\nu_2} = \frac{2R^2}{\nu_2} & \!\! =\!\!  & 39.4\,\,
         \alpha^{-4/5} \eta^{-1} 
         M_{1.4}^{1/4}\dot{M}_{-8}^{-3/10} R_{11}^{5/4} \un{days}{},\\
   \nonumber
   \label{eq:tgamma}
   t_\Gamma  = ~~\frac{R}{\Gamma}~~ &\!\!  =\!\!  & 1.13\,\,
         \alpha^{-4/5} 
	 \left(\frac{\epsilon}{0.1}\right)^{-1} \\
       & &  M_{1.4}^{3/4}\dot{M}_{-8}^{-3/10} R_{11}^{3/4}\un{days}{}, \\
   \nonumber
   t_{\Omega\sub{p}} = \frac{2\pi}{\Omega\sub{p}} &\!\! =\!\! & 
	 7.92\,\,
         \left(\frac{P\sub{orb}}{1\un{day}{}}\right)^2
         \left(\frac{1+q}{q}\right) \times \\
   \label{eq:tprec}
	 & & \hspace{2cm}
         M_{1.4}^{1/2} R_{11}^{-3/2}\un{days}{}.
\end{eqnarray}
Here $\epsilon=L_*/\dot{M}c^2$ measures the radiative efficiency of
the accretion process.  The radiative growth time, $t_\Gamma$, is
especially uncertain because various factors may intervene to make the
radiation absorbed by a disc element different from (and usually less
than) what we would estimate from our direct observations of the X-ray
flux. In particular, the central source may not radiate completely
isotropically. And furthermore, estimating the X-ray flux intercepted
by the disc from its brightness in the UV/optical is complicated by
the fact that only a fraction $1-A$ of the incident flux is
absorbed, where $A$ is the disc albedo.  For X rays incident
perpendicular to a stellar atmosphere, values of $A$ in the range 0.3--0.5
apply, but for the grazing incidence we have here it may be less.
Simultaneous measurements of the X-ray and reprocessed optical fluxes
from X-ray binaries imply very small fractions, $f$, of reprocessed
radiation. For example, the effective temperature of the disc in
Cen\,X-4 implies $f=0.07$ (Heemskerk \& van Paradijs
1989)\nocite{hp:89}. In X-ray bursts with coincident detection of
bursts of reprocessed optical radiation the implied value of $f$ is at
most $10^{-3}$ (see Van Paradijs \& McClintock 1995)\nocite{pm:95}.
It is not likely that such low values are entirely the result of
albedo; possibly, the outer parts of the disc are shadowed from the
central source, due to non-monotonic behaviour of the disc thickness.
Since in warped discs the tilt is usually rather larger than the
expected disc thickness, this effect is probably not so important in
our present considerations.  In all, we cannot expect the measured
precession times to agree with the simple calculations here to better
than a factor few.  Below we shall make the comparison between the
observed long periods and expected radiative periods, where we
estimate the latter as $P_\Gamma= 2\pi t_\Gamma(R\sub{tide})$ in view
of the results from \Sect{numer.pp}.

%-------------------------------------------------------------------------------
      \subsection{System parameters and simulation parameters}
      \label{appli.trans}
%-------------------------------------------------------------------------------

In order to determine which numerical parameters are relevant to a
given binary system we have to know the binary parameters and compute
the appropriate values of various numerical parameters. In particular
we need estimates of the value of $\gamma$ at $R_{\rm tide}$,
$\gamma\sub{tide}$, which is a measure of the strength of the
instability, of $\omega\sub{p,tide}$, which is the relevant
dimensionless measure of the forced precession rate at the outer disc
edge, and of $t_0$ and $\eta$.  In Table~\ref{ta:bin} we list the
basic parameters of a number of X-ray binaries. Most contain neutron
stars, for which a mass of 1.4\msun\ has been assumed. Parameters
derived from the basic ones are given in Table~\ref{ta:der}.
\begin{table}
   \caption{Basic parameters of systems adopted in this study.
	    \label{ta:bin}
	    }
   \begin{tabular}{@{}l|r@{.}lr@{.}lr@{.}lr@{.}lr@{.}l@{}}  \hline
name          & \multicolumn{2}{c}{$P\sub{orb}$}
              & \multicolumn{2}{c}{$M\sub{X}$}
              & \multicolumn{2}{c}{$M\sub{donor}$}
              & \multicolumn{2}{c}{$\dot{M}_{-8}$}
              & \multicolumn{2}{c}{$P\sub{long}$}
              \\
	      & \multicolumn{2}{c}{(days)}
              & \multicolumn{2}{c}{(\msun)}
              & \multicolumn{2}{c}{(\msun)}
              & \multicolumn{2}{c}{}
	      & \multicolumn{2}{c}{(days)}
	      \\ \hline
LMC X-4      &  1&408 &  1&4  & 16&   &  3&   &  30&4 \\ 
Cen X-3      &  2&09  &  1&4  & 18&   &  0&9  & $\sim$140&  \\ 
SS433        & 13&    & 10&   & 20&   & 10&   & 164&  \\ 
X 1907$+$097 &  8&38  &  1&4  & 20&   &  0&05 &  42&  \\ 
LMC X-3      &  1&7   & 10&   &  6&   &  3&   & 198&  \\ 
SMC X-1      &  3&9   &  1&4  & 18&   &  4&   &  $\sim$55&  \\ 
Cyg X-1      &  5&6   & 16&   & 33&   &  2&5  & 294&  \\ 
Her X-1      &  1&7   &  1&4  &  2&35 &  0&25 &  35&  \\ 
X 2127$+$119 &  0&713 &  1&4  &  0&9  &  0&01 &  37&  \\ 
Cyg X-2      &  9&844 &  1&4  &  0&7  &  1&   &  78&  \\ 
gen. LMXB    &  0&2   &  1&4  &  0&5  &  0&1  &   0&  \\ 
X 1916$-$053 &  0&035 &  1&4  &  0&1  &  0&1  &   3&8 \\ 
4U 1626$-$67 &  0&029 &  1&4  &  0&03 &  0&05 &   0&  \\  \hline
\end{tabular}
\end{table}

\begin{table*}
   \caption{Derived parameters using the relations in this paper,
            assuming $\alpha=1$, $\eta=1$, and $\epsilon=0.1$
            (and $M\sub{X}=1.4\msun$). The time scales in the last three 
	    columns are taken at $R\sub{tide}$.
            \label{ta:der}
            }
   \begin{tabular}{@{}l|r@{.}lr@{.}lr@{.}lr@{.}lr@{.}lr@{.}lr@{.}lr@{.}l} \hline
name          & \multicolumn{2}{c}{$q$}
              & \multicolumn{2}{c}{$a_{11}$}
              & \multicolumn{2}{c}{$R\sub{L}/a$}
              & \multicolumn{2}{c}{$R\sub{tide,11}$}
              & \multicolumn{2}{c}{$R\sub{J}/R\sub{tide}$}
              & \multicolumn{2}{c}{$t_{\nu_2}$}
              & \multicolumn{2}{c}{$t_{\Omega\sub{p}}$}
              & \multicolumn{1}{c}{$P_{\Gamma}$} \\
              & \multicolumn{2}{c}{}
              & \multicolumn{2}{c}{}
              & \multicolumn{2}{c}{}
              & \multicolumn{2}{c}{}
              & \multicolumn{2}{c}{}
              & \multicolumn{2}{c}{(days)}
              & \multicolumn{2}{c}{(days)}
              & \multicolumn{1}{c}{(days)} \\ \hline
LMC X-4      & 11&4   &  9&5  & 0&20  &  1&7  & 0&26 &   53&&   8&1 &   7&4 \\ 
Cen X-3      & 12&9   & 13&   & 0&19  &  2&2  & 0&26 &  110&&  12&  &  13&  \\ 
SS433        &  2&    & 50&   & 0&32  & 14&   & 0&25 &  890&& 100&  & 110&  \\ 
X 1907$+$097 & 14&3   & 34&   & 0&19  &  5&4  & 0&26 &  800&&  47&  &  62&  \\ 
LMC X-3      &  0&6   & 11&   & 0&42  &  3&9  & 0&29 &  250&&  21&  &  62&  \\ 
SMC X-1      & 12&9   & 19&   & 0&19  &  3&3  & 0&26 &  110&&  22&  &  11&  \\ 
Cyg X-1      &  2&1   & 34&   & 0&32  &  9&4  & 0&25 &  900&&  44&  & 180&  \\ 
Her X-1      &  1&7   &  6&5  & 0&34  &  1&9  & 0&25 &  130&&  14&  &  17&  \\ 
X 2127$+$119 &  0&64  &  3&1  & 0&42  &  1&1  & 0&29 &  180&&   8&7 &  31&  \\ 
Cyg X-2      &  0&5   & 17&   & 0&44  &  6&6  & 0&30 &  420&& 140&  &  29&  \\ 
gen. LMXB    &  0&36  &  1&2  & 0&47  &  0&51 & 0&33 &   34&&   3&3 &   8&5 \\ 
X 1916$-$053 &  0&07  &  0&36 & 0&60  &  0&19 & 0&50 &   10&&   1&8 &   4&0 \\ 
4U 1626$-$67 &  0&02  &  0&31 & 0&68  &  0&18 & 0&75 &   12&&   4&0 &   4&9 \\ 
 \hline
\end{tabular}
\end{table*}
    
For most of the simulations, we use a disc that extends from $r\sub{in}=1$ to 
$r\sub{tide}=30$.
Since the inner disc does not matter much for our results 
we scale the parameters for each system so
that $R\sub{tide}$ in the system corresponds to the truncation radius in the
numerical grid. Mass is added to the disc at 
$r\sub{J}=r\sub{tide}R\sub{J}/R\sub{tide}$ (see Table~\ref{ta:num}). The
appropriate value of $\omega\sub{p,tide}$ follows from
\begin{eqnarray}
   \nonumber 
   \omega\sub{p,tide} & =&  \Omega\sub{p}(R\sub{tide})R\sub{tide}^2/\nu_1
                            = \frac{\eta}{2} \Omega\sub{p}(R\sub{tide})
			      t_{\nu_2}(R\sub{tide}) \\
   \nonumber
   & = & -15.63~\alpha^{-4/5} M_{1.4}^{-1/4} \dot{M}_{-8}^{-3/10}
		   \left(\frac{P\sub{orb}}{1\un{day}{}}\right)^{-2} \times \\
   \label{eq:omegap}
   &   & \hspace*{2cm}\left(\frac{q}{1+q}\right)
		   R\sub{tide,11}^{11/4}.
\end{eqnarray}
$\omega\sub{p,tide}$ depends on the details of viscosity because it measures the
precession rate, which is independent of viscosity, in units of the viscous
time which is of course viscosity-dependent. In  the numerical grids,
$\omega\sub{p,tide}=\omega\sub{p0}r\sub{tide}^{11/4}$. From the simulations in 
\Sect{numer.res.5}, correcting by a factor 2 for the higher surface densities
implied by the large grids (\Sect{numer.big}), we find that tidal shear
destroys the warp of a disc near the stability limit when $\omega\sub{p0}$
is about $-$0.003, which means $\omega\sub{p,tide}\simeq-35$. This implies 
that most systems have little enough forced precession to have their warps
survive (\Tab{ta:num}), but not by much; also forced
precession may somewhat shorten retrograde long periods
(\Sect{numer.res.2}).

Similarly, $\gamma\sub{tide}$
can be related to the parameters of a real system via
\begin{eqnarray}
   \nonumber
   \gamma\sub{tide} & = & \frac{t_{\nu_2}(R\sub{tide})}{t_\Gamma(R\sub{tide})}
	          \\
   \label{eq:gammatide}
           & = & 0.0287 \left(\frac{\epsilon}{0.1}\right)^{-1}\,\eta
		  M_{1.4}^{1/2} R\sub{tide,11}^{-1/2}
\end{eqnarray}
Note that $\gamma$ does not depend on details of the viscosity. This is
because both the surface density and the viscous time scale are inversely
proportional to the viscosity, so the effect that a lower-viscosity
disc has a longer damping time for disc tilts is exactly compensated
by the fact that its mass also increases the growth time of the
radiative instability.  In table~\ref{ta:num} we list the values of the
numerical parameters that follow from these relations for all systems
in table~\ref{ta:bin}. We also list $t_0$, which is the real time that
passes in each system per unit of dimensionless time in the simulation.
\begin{table}
   \caption{The numerical simulation parameters appropriate to each system
	    in table~\ref{ta:bin} using the same parameters.
	    \label{ta:num}
	    }
   \begin{tabular}{@{}l|r@{.}lr@{.}lr@{.}lr@{.}l@{}} \hline
name         & \multicolumn{2}{c}{$\omega\sub{p,tide}$}
             & \multicolumn{2}{c}{$\gamma\sub{tide}$}
             & \multicolumn{2}{c}{$r\sub{J}$}
             & \multicolumn{2}{c}{$t_0$} \\
	     &   \noval  & \noval&\noval & \multicolumn{2}{c}{(days)} \\ \hline
LMC X-4      & $-$21&  & 0&022 &  7&8 &  0&38 \\ 
Cen X-3      & $-$28&  & 0&020 &  7&9 &  0&75 \\ 
SS433        & $-$27&  & 0&020 &  7&5 &  6&3  \\ 
X 1907$+$097 & $-$54&  & 0&012 &  7&9 &  5&7  \\ 
LMC X-3      & $-$37&  & 0&039 &  8&8 &  1&8  \\ 
SMC X-1      & $-$16&  & 0&016 &  7&9 &  0&81 \\ 
Cyg X-1      & $-$65&  & 0&032 &  7&5 &  6&4  \\ 
Her X-1      & $-$30&  & 0&021 &  7&6 &  0&94 \\ 
X 2127$+$119 & $-$66&  & 0&027 &  8&6 &  1&3  \\ 
Cyg X-2      &  $-$9&6 & 0&011 &  9&1 &  3&0  \\ 
gen. LMXB    & $-$32&  & 0&040 &  9&8 &  0&24 \\ 
X 1916$-$053 & $-$17&  & 0&066 & 15&  &  0&07 \\ 
4U 1626$-$67 &  $-$9&2 & 0&067 & 22&  &  0&08 \\  \hline
\end{tabular}
\end{table}

%-------------------------------------------------------------------------------
      \subsection{Comparing forced and radiative precession with data}
      \label{appli.compa}
%-------------------------------------------------------------------------------

Let us now assume that the disc radii in real binaries are adequately
approximated by $R\sub{tide}$, and that the disc tilt is big enough that
mass input occurs predominantly at $R\sub{J}$. Then we can predict what
the forced and radiative precession time scales are for all the systems in 
table~\ref{ta:num} and compare them with their observed long periods.

\begin{figure}
\epsfxsize=\columnwidth\epsfbox{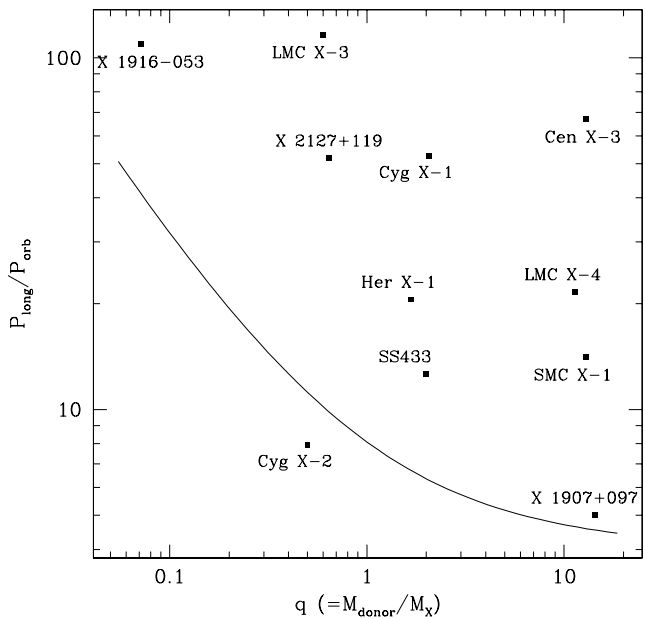}\hfill%
   \caption{The observed long periods as a function of mass ratio. The solid
            curve indicates the expected forced precession period.
	    \label{fi:forceprec}
	   }
\end{figure}

For forced precession, a particularly simple relation ensues if we
divide the result by the orbital period: \be{eq:forceprecq}
\frac{P\sub{p}}{P\sub{orb}} = \frac{4}{3}
\left(\frac{0.87R\sub{L}}{a}\right)^{-3/2}\frac{\sqrt{1+q}}{q} \ee
Since $R\sub{L}/a$ is only a function of mass ratio, so is the whole
right-hand side. This presents us with a natural ordering of the
systems by mass ratio, which is shown in Fig~\ref{fi:forceprec}. The
curve predicted by \Eq{eq:forceprecq} is also shown, and while it is
within an order of magnitude of the data, it does not appear to
predict the observed values well at all. Specifically, even an
arbitrary vertical shift of the predicted line cannot fit the data
well, simply because no trend of the observed long periods with mass ratio is
evident in the data, and both at mass ratio unity and at mass ratio 10 the
observed values range over a factor 15. Thus a reasonable fit can be
obtained only by omitting some of the data points (Larwood 
1998)\nocite{larwo:98}.

\begin{figure}
\epsfxsize=\columnwidth\epsfbox{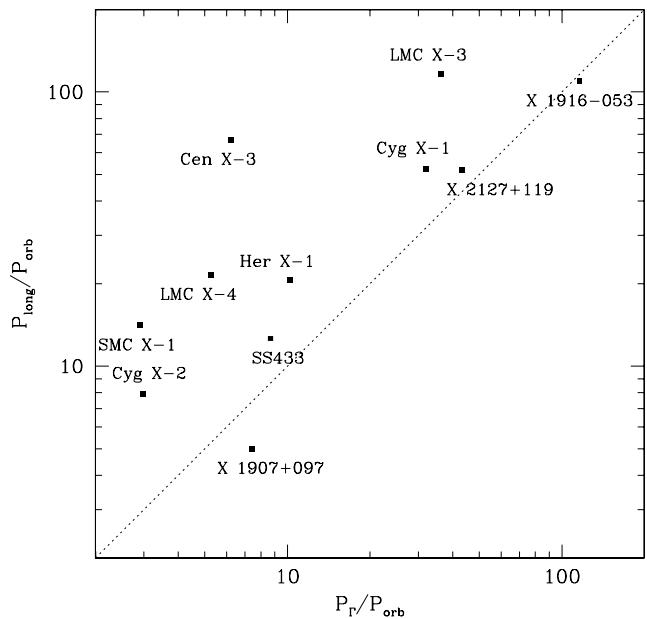}
   \caption{Comparison
	    of the observed long periods with the expected radiative
	    precession periods. The dashed line indicates equality.
	    \label{fi:radprec}
	   }
\end{figure}

When we repeat the exercise to test the radiative precession model there
is not one simple system parameter with which the times scale, so we plot
the same ordinate as in the previous figure versus the predicted
$P_\Gamma/P\sub{orb}$; a rather different picture results
(\Fig{fi:radprec}).  There is a reasonable correlation between predicted
and observed periods over the range of a factor 40 in
$P_\Gamma/P\sub{orb}$, with a mean ratio somewhat below unity (note that there
was no adjustment of the normalisation). Since there is some freedom in
the choice of disc albedo and $\alpha$ parameter, we can adjust the mean
ratio of predicted to observed period to be one, e.g.\ by setting $\alpha=0.27$
(\Sect{appli.herx1}). Then most systems come within a factor 2 of the curve,
with the exception of Cen\,X-3 and X\,1907$+$097, which for very similar
predicted periods have a factor 14 difference in observed long period.
(Note that they also have almost the same mass ratio, so forced precession
faces the same problem of a large difference between systems predicted to be
the same.) On the general ability to predict long periods, the
radiative precession model therefore does much better than tidally forced
precession. It is perhaps remarkable that SS\,433, which is said to have a
thick disc and for which our model should therefore not be valid, fits the
relation rather well. One might wish to consider the possibility that the
(outer) disc of SS\,433 subtends a large solid angle not because it is
intrinsically thick, but because it is strongly warped and tilted.

\begin{figure}
\epsfxsize=\columnwidth\epsfbox{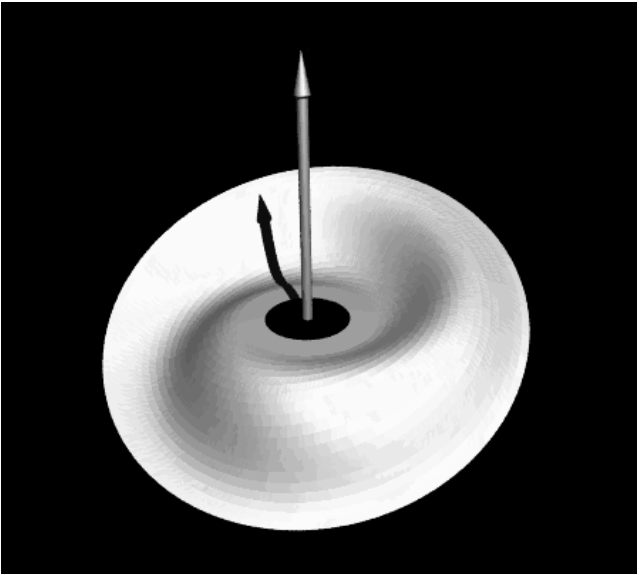}
   \caption{The shape of a disc with mass input well inside the outer
            edge. It flares to high tilts near the outer edge and precesses
	    retrogradely under the action of radiation torques.
	    (run with small grid, $r\sub{add}=10$, $r\sub{tide}=30$,
	    $F_\star=0.045$)
	    \label{fi:retrdisc}
	   }
\end{figure}

\begin{figure}
\epsfxsize=\columnwidth\epsfbox{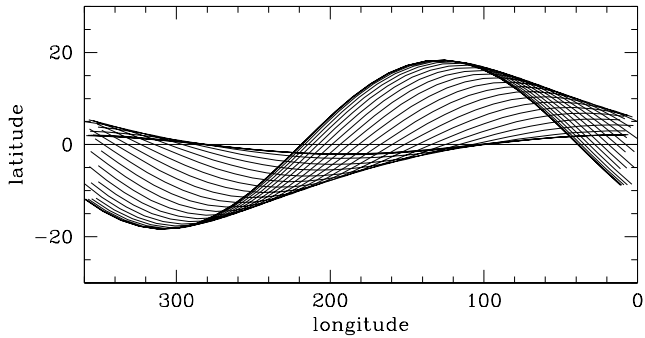}
   \caption{A projection of the disc shape for a mildly inclined retrogradely
	    precessing disc onto the sky as seen from the central source.
	    (run with small grid, $r\sub{add}=10$, $r\sub{tide}=30$,
	    $F_\star=0.045$)
	    \label{fi:skyretrdisc}
	   }
\end{figure}

\begin{figure}
\epsfxsize=\columnwidth\epsfbox{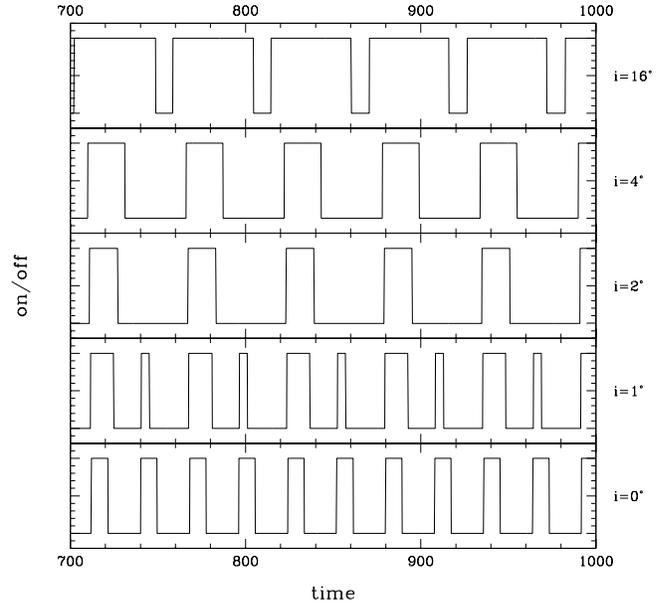}
   \caption{Curves marking X-ray on and off states for the precessing disc
            of \Fig{fi:skyretrdisc} as seen for different inclinations of the
	    binary orbit to the line of sight. For high inclinations only
	    one on state occurs per period, but for $i=1^\circ$ one observes
	    an alteration of long-on and short-on states, similar perhaps
	    to the main-high and short-high states of Her\,X-1
	    (Jones \& Forman 1976)\protect\nocite{jf:76}.
	    (run with small grid, $r\sub{add}=10$, $r\sub{tide}=30$,
	    $F_\star=0.045$)
	    \label{fi:xonoffretr}
	    }
\end{figure}

Our model can be tested further by probing directly the shapes of the
discs using eclipse mapping and reverberation mapping. In
\Fig{fi:retrdisc} we show the shape of the disc as seen in a
simulation of a fixed-shape retrogradely precessing outer disc that might
correspond to the case of Her\,X-1. Note the flaring towards the
outer edge and the near alignment with the orbital plane inside the
mass input radius. We remind the reader that the range of luminosities
over which these stationary solutions exist is quite narrow (We found
it to be  only 30\%). But whether Her\,X-1 has a constant
tilt is not really known, though large variations in the outer-disc
tilt appear not to be needed to explain the photometry. However, the
source is known to have anomalous on- and off states or dips (Giacconi
et~al.\ 1973)\nocite{ggkls:73}, which could well be caused by a
variation of the disc tilt. In \Fig{fi:skyretrdisc} we show a
projection of the disc onto the sky as seen from the central
source. An observer at infinity would be represented by a straight
line at constant latitude as the disc precesses. When the line is
clear of the disc, which can happen either once or twice per
precession period, the observer sees an on state of the X-ray source.
For low inclinations, one sees two on states of different length
(\Fig{fi:xonoffretr}), once when the line of sight passes under the
disc (short-on) and once when it passes over it (long-on).

\begin{figure}
\epsfxsize=\columnwidth\epsfbox{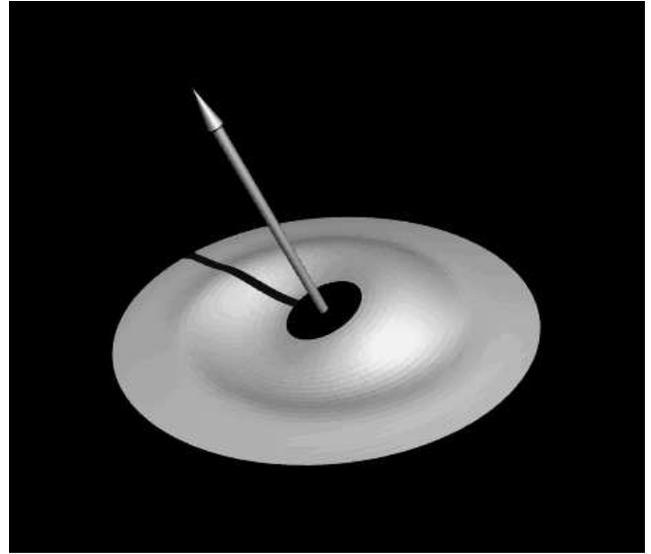}
   \caption{The shape of a disc with mass input close to the outer
            edge. Its highest tilt occurs at the inside and the bulk of
	    it precesses
	    progradely under the action of radiation torques.
	    (run with small grid, $r\sub{add}=20$, $r\sub{tide}=30$,
	    $F_\star=0.2$)
	    \label{fi:progdisc}
	   }
\end{figure}

\begin{figure}
\epsfxsize=\columnwidth\epsfbox{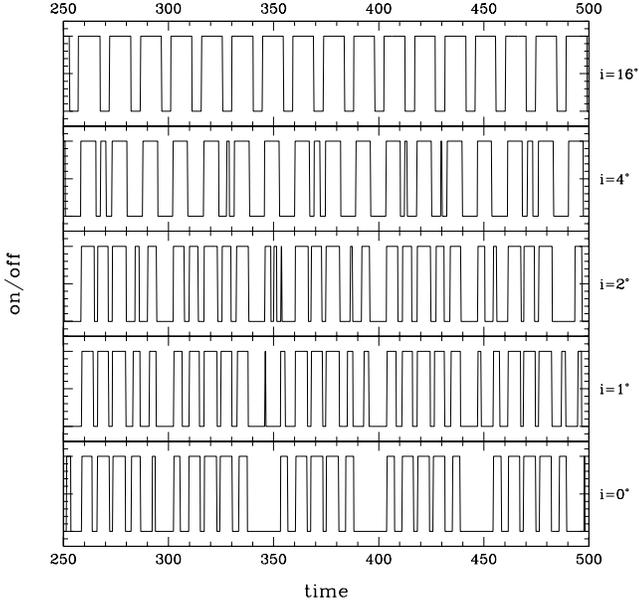}
   \caption{Curves marking X-ray on and off states for the precessing disc
            of \Fig{fi:progdisc} as seen for different inclinations of the
	    binary orbit to the line of sight. For high inclinations only
	    one periodicity is visible, but for lower ones the beat between
	    the outer disc and inner disc periods is clearly visible.
	    (run with small grid, $r\sub{add}=20$, $r\sub{tide}=30$,
	    $F_\star=0.2$)
	    \label{fi:xonoffprog}
	   }
\end{figure}

By contrast,
a progradely precessing disc such as perhaps that of Cyg\,X-2 has
its highest tilt in the middle. A typical example is shown in
\Fig{fi:progdisc}. When mass input in such a disc is not at the very outside
edge the outer part can be going in a retrograde direction at the same time,
with a different period (\Sect{numer.res.4}). Under favourable inclinations
an observer could see both the outer and inner disc periodicities in the
sequence of X-ray on and off states. An example of this is shown in
\Fig{fi:xonoffprog}.

%-------------------------------------------------------------------------------
      \subsection{Hercules\,X-1}
      \label{appli.herx1}
%-------------------------------------------------------------------------------

For the specific case of Her\,X-1, where a relatively stable cycle exists
over a long time, and the retrograde precession has been established well
from observations, we may look at the numbers in somewhat greater detail.
Using $\gamma\sub{crit}=0.1$ and \Eq{eq:tnu2}, we can establish a 
maximum possible radiative precession time scale
\begin{eqnarray}
  \nonumber 
  P\sub{p,max} & = & 2\pi\gamma\sub{crit}t_{\nu_2} \\
  \label{eq:herx1pmax}
               & = & 82.5 \alpha^{-4/5}\eta^{-1} \un{days}{},
\end{eqnarray}
where we have used the parameters of Her\,X-1 from our tables. The
stable long period indicates that Her\,X-1 must be within 30\% of the
stability boundary (\Sect{numer}), so supposing we equate this period
to the actually observed value of 35\,d, this gives \be{eq:herx1alpha}
\alpha^{4/5}\eta = 2.36.  \ee If we then use $\eta=1/(2\alpha^2)$
(Kumar and Pringle 1985, Ogilvie 1998)\nocite{kp:85,ogilv:98}
we find $\alpha=0.27$, which is not an unreasonable value (see
Section 5). Note that our results imply quite generally that the
measurement of a precession period constitutes a crude measurement of
the disc viscosity.

Rather than assuming near-criticality, we may also just compute $P_\Gamma=2\pi
t_\Gamma$ from \Eq{eq:tgamma} to get
\be{eq:herx1pgamma}
   P_\Gamma = 17.3 \alpha^{-4/5}\frac{0.1}{\epsilon},
\ee
which shows good agreement with the observed 35-d period for reasonable values
of $\alpha$ and $\epsilon$.

%-------------------------------------------------------------------------------
      \subsection{Very long periods: 4U\,1820$-$30 and X\,1916$-$053}
      \label{appli.verylong}
%-------------------------------------------------------------------------------

Two well-documented long periods which are not shown in Figs.~\ref{fi:forceprec}
and \ref{fi:radprec} are the 176-day period of 4U\,1820$-$30 and the
199-day period of X\,1916$-$053 (Priedhorsky \& Terrell 1984b)\nocite{pt2:84}.
The reason is that both would be very
far off the top of the vertical axis: these periods are too long in such
compact binaries to be a disc precession period. For 4U\,1820$-$30 the
bursting behaviour changes with the 176-day cycle, suggesting that indeed
the accretion rate varies though the cycle rather than just the aspect of
a disc. In X\,1916$-$053 a third  star orbiting the binary has been
advanced as the cause of the very long period (Grindlay
et~al.\ 1988\nocite{gbclt:88}).

%-------------------------------------------------------------------------------
      \subsection{Torque reversals}
      \label{appli.torq}
%-------------------------------------------------------------------------------

Some X-ray pulsars with long and well-sampled pulse period records show
very extended epochs in which the spin period decreases, followed by
equally long spindown epochs. A famous example is 4U\,1626$-$67 (Nelson
et~al.\ 1997, Chakrabarty et~al.\ 1997)\nocite{nbcfk:97,cbgkp:97}, and
another is GX\,1$+$4. We note that in our simulations many of the more
luminous discs are tilted through more than 90 degrees in their inner
parts (Figure~\ref{fi:bigtiltfstarin}), implying that they are rotating
counter to the usual direction and would provide a torque that is always
spindown, assuming the pulsar spin is prograde relative to the orbit. The
outer disc, at the same time, can be tilted much less strongly
(Figure~\ref{fi:bigtiltfstar}). Since the range in luminosity over which
the discs are unstable but only have small tilt is fairly small, and since
all X-ray binaries show long-term flux variations, it is quite possible
that a few binaries have luminosities that vary around the value that
separates prograde and retrograde inner discs.
More detailed discussion of this torque reversal model is given in a
companion paper (van Kerkwijk et~al.\ 1998)\nocite{kcpw:98}.

%%%%%%%%%%%%%%%%%%%%%%%%%%%%%%%%%%%%%%%%%%%%%%%%%%%%%%%%%%%%%%%%%%%%%%%%%%%%%%%%
   \section{Discussion}
   \label{discu}
%%%%%%%%%%%%%%%%%%%%%%%%%%%%%%%%%%%%%%%%%%%%%%%%%%%%%%%%%%%%%%%%%%%%%%%%%%%%%%%%

%-------------------------------------------------------------------------------
      \subsection{Warp wave propagation}
      \label{discu.waves}
%-------------------------------------------------------------------------------

	The equation we use to describe the time-evolution of the disc
tilt makes the basic assumption that angular momentum is transferred
between disc elements predominantly via viscous processes (Pringle
1992)\nocite{pring:92}. 
However, fluid discs can also propagate angular momentum, and
therefore warps, through wave-like processes (Lubow \& Pringle 1993,
Korycansky \& Pringle 1995, Papaloizou \& Lin 1995, Lubow \&
Ogilvie 1998)\nocite{lp:93,kp:95,pl:95,lo:98}.
For wave-like propagation to dominate, it is necessary
that the angular velocity in the disc be closely Keplerian, and for
the viscosity to be sufficiently small (see the discussion in Pringle
1997)\nocite{pring:97}. By closeness to Keplerian we mean that
\be{eq:wavelimit1}
    \frac{|\Omega-\kappa|}{\Omega} \ll \frac{H}{R},
\ee
throughout the disc, where $\kappa$ is the epicyclic frequency.
Making the approximation that the binary
companion can be represented as a ring of mass $M_2$ at distance $a$
from the primary, mass $ M_1$, we find that, to a first approximation in
($R/a$), 
\be{eq:kappa2}
    \frac{|\Omega-\kappa|}{\Omega} = \frac{5q}{8} \left(\frac{R}{a}\right)^3.
\ee
Taking the system parameters for Her\,X-1 in Tables 2 and 3 we find
that near the edge of the disc, $R = R\sub{tide}$,
\be{eq:kappa3}
    \frac{|\Omega-\kappa|}{\Omega} = 
           0.0261 \left(\frac{R}{R\sub{tide}}\right)^3.
\ee
By comparison we find that
\be{eq:hoverr}
	\frac{H}{R} = 0.04 \alpha^{-1/10} \dot{M}_{-8}^{-3/20} M_{1.4}^{-3/8} 
	      \left(\frac{R}{R\sub{tide}}\right)^{1/8}.
\ee
Thus close to the disc edge we expect wave propagation to be
marginally possible by this criterion, and to improve significantly at
smaller radii.

	However, for wave propagation we also require that the disc
viscosity be sufficiently small. A disc warp wave in a Keplerian disc
propagates a distance of order $H/\alpha$, before it is dissipated by
viscosity. Thus for a wave to be able to propagate over a radial
distance R, we require
\be{eq:wavelimit2}
	\alpha \ll \frac{H}{R}.
\ee
Measurement of $\alpha$ in the discs in the binary systems we are
discussing is not easy, since the magnitude of the viscosity is best
estimated from measuring the time scale on which the disc as a whole
varies. However, good estimates for $\alpha$ are available for a related
set of binary systems, the X-ray novae. For these systems it is
possible to parametrise the quantity $\alpha$ by modelling the X-ray
light curves through the outburst cycle (mainly the decline from
outburst), and the result obtained, which it turns out also fits the
outbursts of dwarf nova systems, is (Cannizzo, Chen \& Livio,
1995)\nocite{ccl:95}
\be{eq:alphaKCL}
	\alpha = 50 \left(\frac{H}{R}\right)^{1.5}.
\ee
For the value of $H/R$ appropriate here (eq.~\ref{eq:hoverr}), this corresponds
to
\be{eq:alphares}
	\alpha = 0.45  \dot{M}_{-8}^{-9/46} 
	         M_{1.4}^{-45/92} \left(\frac{R}{R\sub{tide}}\right)^{15/92},
\ee
which is also in line with the values of $\alpha = 0.1 - 1.0$ required
for discs in dwarf nova outbursts (e.g.\ Cordova
1995)\nocite{cordo:95}, and for the value of $\alpha$ which we require
to bring our calculations into agreement with the parameters of Her
X-1 (Section 4.4).

Thus wave propagation can occur in these discs only if \be{eq:hrlimit}
\frac{H}{R} \ll 0.01 \dot{M}_{-8}^{-1/11} M_{1.4}^{-5/22}
\left(\frac{R}{R\sub{tide}}\right)^{5/66}.  \ee We conclude that, in
contrast to the calculations of Larwood (1998), the discs in binary
X-ray systems are unlikely to be able to sustain propagating warp
waves, and that the diffusion based approach is likely to provide a
better description of the physics of warp evolution in these systems.

%-------------------------------------------------------------------------------
      \subsection{Possible consequences of some neglected effects}
      \label{discu.neglect}
%-------------------------------------------------------------------------------

In all the simulations we used the time-independent expression for the
forced precession, i.e.\ we assumed the precession period to be much
longer than the orbital period. In reality $P\sub{long}/P\sub{orb}$ is
typically a few tens, but can be as low as 5. This means deviations
from the smooth precession and extra nodding motions of the disc with
periods that are beats between multiples of the precession and orbital
periods can occur. These have in fact been seen in SS\,433 and
discussed in these terms (Katz et~al.\ 1982)\nocite{kamg:82}. In
addition to this short-period effect, mass input would also be
influenced by the varying phase between companion star and disc: the
accretion stream would first hit the disc somewhere along its curved
intersection with the orbital plane, and the intersection radius would
vary between the outer disc edge and the minimum distance of approach
to the centre allowed by the stream's angular momentum (about
0.5$R\sub{J}$). So a more realistic simulation would either average
the mass input over a large range in disc radii or have a
time-variable mass input radius. Perhaps such effects could also
explain the curious fact that in at least some systems (Her\,X-1,
LMC\,X-4, and Cyg\,X-2) the precession period is an integer multiple
of the orbital period.

The luminosity we used in any simulation was strictly constant. For all
stationary solutions this should be valid, but the solutions in which the
outer disc tilt is large and variable (i.e.\ at high luminosity; 
\Sect{numer.big}) are not strictly valid: their surface densities are
rather lower when the tilt is large than when it is small, 
and since mass is added
to the disc at a constant rate this must mean that the amount of mass passing
through the inner disc edge varies with time, and therefore the central source
luminosity should do so too. This extra feedback operates with a delay
of order the outer disc viscous time, and such delayed feedback could
lead to further instabilities (as any thermostat designer knows).

Schandl \& Meyer (1994)\nocite{sm:94} and Schandl
(1996)\nocite{schan:96} have studied the effects of a disc emitting
wind.  Like radiation the wind carries momentum and therefore its
launch from the disc surface causes a force on the disc. Since the
momentum per unit energy in a slow wind is much greater than in
radiation this could be a more important effect. However, we
understand the formation of winds only poorly and therefore the { \it
ab initio } calculation of how much wind there is and how much
momentum it carries away is difficult. Unfortunately, Schandl \& Meyer
neglected self-shadowing and used an older and incorrect equation of
motion for the disc, so it is difficult to make a detailed comparison
of the two models and the reasons why they differ. The predicted disc
shapes are strikingly different though, so the issue may well be
observationally decidable: their discs have tightly wound spirals with
the angle between the line of nodes and the radial direction
decreasing outward, whereas the radiative warps have very open
spirals, and the angle of the line of nodes with the radial direction
increases outwards. If the sonic point of the wind were close to the
disc, unlike in the study by Schandl \& Meyer, then the behaviour
would be almost the same as for our radiation-driven precession,
except that the force could be much greater for a given incident
flux. This means that the region of instability moves to lower
luminosities. Since value of $\gamma\sub{crit}$ we find for our model
is not too far from the actual values in X-ray binaries, this would
mean that wind-driven warping would give extremely unstable, erratic
behaviour in those systems, which appears to contradict the regular
behaviour of Her\,X-1.

%%%%%%%%%%%%%%%%%%%%%%%%%%%%%%%%%%%%%%%%%%%%%%%%%%%%%%%%%%%%%%%%%%%%%%%%%%%%%%%%
   \section{Conclusion}
   \label{conclu}
%%%%%%%%%%%%%%%%%%%%%%%%%%%%%%%%%%%%%%%%%%%%%%%%%%%%%%%%%%%%%%%%%%%%%%%%%%%%%%%%

We have examined the viability of radiation-driven warping and
precession of thin accretion discs as a model for the long or third
periods in X-ray binaries. The model holds up very well and has a
number of advantages over earlier proposals. (i) It not only causes a
disc to precess once tilted, but also gives rise to and maintains the
tilt, whereas in previous models the origin of the tilt was always a
difficult issue. (ii) The equation of motion for the disc we use does
not suffer from the physical inconsistencies of some earlier proposed
equations of motion. Also, unlike linearised approximations to the
solutions of our equation in earlier papers and some other studies,
this numerical study takes account of the important effects of
self-shadowing of the disc. (iii) The quantitative agreement between
the observed long periods of X-ray binaries and the results of our
simulations is good, whereas forced precession due to the companion's
gravitational pull on the disc fares rather poorly when compared
quantitatively with the data. Also, radiative precession can be
prograde as well as retrograde, unlike forced precession. There is
tentative evidence that the discs in Cyg\,X-2 and possibly
X\,1916$-$053 do precess progradely. To put it succinctly, we have
shown that if the radiative instability of Pringle (1996) gives rise
to the disc tilt in these systems, then it also automatically gives
rise to disc precession at approximately the observed rate. Moreover,
we have also shown that if tidally induced precession becomes
dominant, then the instability is likely to be stabilised, and the
disc to remain in the orbital plane.

In addition to steadily precessing discs, radiative warps at
high luminosity can also be non-stationary: their tilt angle varies
periodically with time in our simulations, and in realistic cases with
feedback between central source luminosity and accretion rate would
probably exhibit non-periodic behaviour as well. This may be applicable
to many of the long X-ray periods observed in nature, which are not very
stable in amplitude and/or period.

One particular feature of high-luminosity systems is that the inner disc may
tilt through more than 90 degrees, and thus rotate counter to the normal
direction. When it encounters the magetosphere of a neutron star it will
then provide a strong spin-down torque, possibly explaining the torque
reversals seen in systems such as 4U\,1626$-$67. One would expect the
X-ray source to be behind the disc much of the time when the warp is so
strong, but the strongly warped discs have much lower surface densities, so
they could be (partly) transparent to X rays (van Kerkwijk et~al.\
1998)\nocite{kcpw:98}.

{\it Acknowledgements\/} RAMJW gratefully acknowledges support from the
Royal Society through a URF grant. We also thank M. Begelman and
P. Maloney for useful discussions, and for a preprint of their paper.

%%%%%%%%%%%%%%%%%%%%%%%%%%%%%%%%%%%%%%%%%%%%%
%  tail stuff: bibliography
%%%%%%%%%%%%%%%%%%%%%%%%%%%%%%%%%%%%%%%%%%%%%
%\bibliographystyle{astroshortmn}
%\bibliography{xmymoshort,x65,x70,x75,x80,x85,x90,x95,x00,xnew,discprec}

\end{document}